%% file: dm_v2submit.tex
\documentclass[
 reprint,preprintnumbers,
 amsmath,amssymb,
 aps,prd,superscriptaddress,
]{revtex4-1}

\usepackage{graphicx}
\usepackage{dcolumn}
\usepackage{bm}
\usepackage{hyperref}
\usepackage[mathlines]{lineno}
\usepackage{placeins}
\usepackage[export]{adjustbox}
\usepackage{ amssymb }
\usepackage{rotating}
\usepackage{lineno}
\usepackage{tabularx}
\usepackage[utf8]{inputenc}
\usepackage[T1]{fontenc}
\usepackage{tikz}
\usepackage{color}

\hypersetup{
    pdfnewwindow=true,    
    colorlinks=true,      
    linkcolor=blue,       
    citecolor=blue,       
    filecolor=blue,       
    urlcolor=blue         
}
\usepackage[normalem]{ulem}

\begin{document}

\preprint{MIT-CTP/5038}

\title{Constraints on Spin-Dependent Dark Matter Scattering with Long-Lived Mediators from TeV Observations of the Sun with HAWC}

\input{authorlist.tex}
\collaboration{HAWC Collaboration\vspace{-0.3cm}}
\input{guest_author.tex}

\date{\today}

\begin{abstract}
We analyze the Sun as a source for the indirect detection of dark matter through a search for gamma rays from the solar disk.  Capture of dark matter by elastic interactions with the solar nuclei followed by annihilation to long-lived mediators can produce a detectable gamma-ray flux. We search three years of data from the High Altitude Water Cherenkov (HAWC) observatory and find no statistically significant detection of TeV gamma-ray emission from the Sun. Using this, we constrain the spin-dependent elastic scattering cross section of dark matter with protons for dark matter masses above 1 TeV, assuming a sufficiently long-lived mediator. The results complement constraints obtained from Fermi-LAT observations of the Sun and together cover WIMP masses between 4 and $10^6$ GeV. In the optimal scenario, the cross section constraints for mediator decays to gamma rays can be as strong as $\sim10^{-45}$~cm$^{-2}$, which is more than four orders of magnitude stronger than current direct-detection experiments for 1 TeV dark matter mass. The cross-section constraints at higher masses are even better, nearly 7 orders of magnitude better than the current direct-detection constraints for 100 TeV dark matter mass. This demonstration of sensitivity encourages detailed development of theoretical models in light of these powerful new constraints.

\end{abstract}

\maketitle


\section{\label{sec:level1}Introduction}
A variety of astrophysical observations, including galaxy rotation curves, large scale structure and cosmic microwave background (CMB) measurements, point towards the existence of non-baryonic dark matter in the Universe \cite{Rubin:1978kmz,Ade:2015xua,doi:10.1146/annurev-astro-082708-101659,2017arXiv171206615B,2016ChPhC..40j0001P,2017IJMPD..2630012F,2011AdAst2011E...8G}. Testing the particle nature of dark matter  candidates through their interactions with baryonic matter is a key aspect of research in physics beyond the Standard Model (SM). 

The scattering cross section of weakly interacting massive particle (WIMP) dark matter can be studied in astrophysical environments of high matter density, such as the Sun. WIMPs from the galactic dark matter halo can be gravitationally trapped by the Sun through scattering off solar nuclei, and settle into thermal equilibrium at the core \cite{PhysRevLett.55.257,2009PhRvD..79j3531P,2017JCAP...05..046W,PhysRevLett.114.081302,1985ApJ...296..679P,Danninger:2014xza}. The overdensity of dark matter in the core can result in the annihilation of dark matter into SM particles \cite{1995NuPhS..43..265E, 2009PhRvD..79j3532P,Bell:2011sn,Feng:2016ijc,Kouvaris:2016ltf}. Once equilibrium has been reached, the flux of the annihilation products only depends on the capture rate, and therefore, the scattering cross section (see Sec. \ref{sec:dm_sun}). 

If dark matter has only spin-dependent elastic scattering interactions, the best sensitivity from direct-detection experiments  \cite{PhysRevLett.118.021303,2017NatPh..13..212L,2016JPhG...43a3001M,PhysRevLett.119.181302} is several orders of magnitude weaker than for spin-independent scattering \cite{2017PhRvL.118g1301F,2017PhRvL.118y1301A,1681014,Akerib:2017kat,PhysRevLett.119.181301, Aprile:2018dbl}. For studying spin-dependent cross sections, indirect methods based on WIMP capture in the Sun (with abundant hydrogen targets) can be substantially more sensitive than direct-detection techniques \cite{Indirectdm,Leane:2018kjk}. IceCube \cite{2017EPJC...77..146A}, ANTARES \cite{2016PhLB..759...69A} and Super-K \cite{Choi:2015ara} have performed searches for the neutrino signatures of annihilating dark matter in the Sun, and constrained the cross sections up to an order of magnitude better than direct-detection experiments for dark matter masses above a few hundred GeV. 

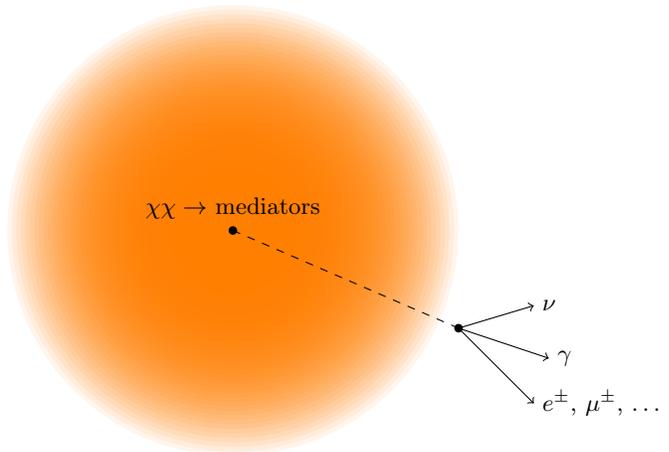
\begin{figure}[ht]

\centering
\input{mediator}

 \caption{Illustration of dark matter annihilation into long-lived
  mediators that decay to SM particles outside the
  solar surface (adapted from Ref. \cite{2017PhRvD..95l3016L}). Dark matter annihilates to long-lived mediators Y via the process DM + DM $\rightarrow$ Y + Y $\rightarrow$ 2 (SM + SM), where each mediator decays to two SM particles. Gamma rays are obtained either from direct mediator decay, hadronic cascades, or bremsstrahlung (see text for details).}
  \label{fig:dm_annihilation}
\end{figure}


If WIMP annihilations produce standard-model particles, then those include gamma rays, produced through decays, bremsstrahlung, or (even for annihilations that proceed only to neutrinos) weak bremsstrahlung. However, gamma rays produced inside the Sun are extinguished by solar matter. In typical WIMP scenarios, the probability of observing a gamma-ray signal from the Sun is extremely low. The thermalized dark matter profile is peaked at the Sun's core, with a very small annihilation rate outside the solar atmosphere \cite{2001hep.ph....3277H,2010PhRvD..81f3502S,2009PhRvD..79j3532P,1997PhDT.........5E}. Such scenarios do not produce a high enough gamma-ray flux that could be probed with ground or satellite-based detectors, as shown in Ref. \cite{2010PhRvD..81f3502S}. 

A different scenario --- with enhanced prospects of gamma-ray detection --- comes from models in which dark matter annihilates into a long-lived mediator that could escape and decay outside the Sun to produce gamma rays, electrons or other SM particles \cite{2017PhRvD..95l3016L, 2010PhRvD..82k5012S,2010PhRvD..81a6002S,2010PhRvD..81g5004B,Curtin:2018mvb,Holdom:1985ag,1998pesu.book....1M, Pospelov:2007mp,Pospelov:2008jd, Batell:2009zp,Rothstein:2009pm,Chen:2009ab,Meade:2009mu,Bell:2011sn,Feng:2015hja,Kouvaris:2016ltf,Feng:2016ijc,Kouvaris:2016ltf,Bell:2016fqf, Bell:2016uhg,Adrian-Martinez:2016ujo,Allahverdi:2016fvl, Ardid:2017lry,Ajello:2011dq,Arina:2017sng}, as illustrated in Fig. \ref{fig:dm_annihilation} and detailed further in Sec. \ref{sec:dm_sun}. A fairly minimal dark sector contains a dark matter candidate, along with a mediator, which allows interaction between the dark and SM sectors. Dark mediators appear naturally in many ultraviolet complete theories, and include examples such as dark photons, dark Higgs, and axions \cite{2010PhRvD..81g5004B,Curtin:2018mvb,Holdom:1985ag,1998pesu.book....1M,Nomura:2008ru,Profumo:2017obk}. If the mediators are light or have small couplings, they can be long-lived, and can decay outside the Sun into detectable gamma rays. 

The prospects for detecting TeV signals from the decay of long-lived mediators outside the Sun with HAWC were first studied in Refs. \cite{2017PhRvD..95l3016L,Arina:2017sng}. It was predicted that the solar gamma-ray channel can provide very strong sensitivity to the dark matter scattering cross sections in the spin-dependent parameter space. In this work, we follow up with observations of the TeV Sun. The High Altitude Water Cherenkov (HAWC) Observatory can search for gamma rays from the Sun in an energy range that was not accessible before. We discuss the analysis and the resulting constraints on gamma rays above 1 TeV obtained by HAWC in a companion paper \cite{Astropaper}. Our search for gamma rays from the Sun falls within an active part of solar cycle 24 (2014--2017) which is important for dark matter searches from the Sun, as described in Sec. \ref{sec:hawc}. 

The paper is structured as follows. Section \ref{sec:dm_sun} outlines the mechanism of dark matter scattering and annihilation in the Sun. Section \ref{sec:hawc} reviews the search for GeV--TeV gamma rays from the Sun and describes the HAWC detector. In Section \ref{constraints}, we calculate the constraints on spin-dependent scattering for various annihilation channels, providing strong new limits. Section \ref{sec:conc} concludes the paper.


\section{\label{sec:dm_sun}Dark Matter in the Sun}
We briefly review the process of WIMP capture and annihilation in the Sun. WIMPs can lose kinetic energy via scattering and settle into thermal equilibrium in the core of the Sun \cite{PhysRevLett.55.257,1995NuPhS..43..265E,2004PhRvD..69l3505L,2009arXiv0908.0899F,2009PhRvD..79j3531P,2017JCAP...05..046W,Feng:2016ijc,Griest:1986yu,Gould:1987ww}. The overdensity of dark matter in the core can result in dark matter annihilation into SM particles. Evaporation is not important for dark matter masses above a few GeV \cite{GRIEST1987681,Gould:1987ir}. Ignoring self-interactions \cite{Zentner:2009is}, the number of dark matter particles $N$ in the Sun, at a time $t$, can be written as a function of the capture and annihilation rates \cite{2017JCAP...05..046W,2017PhRvD..95l3016L},

\begin{equation}
\frac{dN}{dt} = \Gamma_{\text{cap}} - C_{\text{ann}}N^2,
\label{eq:capann}
\end{equation}
where $\Gamma_{\text{cap}}$ is the capture rate, and $C_{\text{ann}}$ is a factor accounting for the annihilation cross section and the dark matter number density. Initially, when the Sun was formed, the capture rate far exceeded the number of annihilation events per unit time, $\Gamma_{\text{ann}}$. Eventually, when capture and annihilation reach equilibrium (dN/dt = 0), the annihilation rate becomes,
\begin{equation}
\Gamma_{\text{ann}} = \frac{1}{2}C_{\text{ann}}N^2 = \frac{1}{2}\Gamma_{\text{cap}}  .
\label{eq:equib}
\end{equation}
The factor of 1/2 accounts for two dark matter particles being depleted in each annihilation event. The annihilation rate in equilibrium is independent of the annihilation cross section $\langle \sigma_Av\rangle$, and is set by $\Gamma_{\text{cap}}$, which depends on the scattering cross section and the local halo mass density, among other things \cite{1997PhDT.........5E,Rott:2011fh}. Observed signals of annihilation would be a direct probe of the WIMP capture rate and therefore, the spin-dependent cross section $\sigma^{\text{SD}}$ \cite{2017PhRvD..95l3016L,2009PhRvD..79j3532P,Gould:1991hx}. In addition, it may be possible to determine the WIMP mass $m_{\chi}$ through a cutoff in the spectrum of its annihilation products. The angular profile of the region where annihilation is concentrated is narrow and embedded deep within the Sun \cite{1997PhDT.........5E}.

Detecting a dark matter signal in gamma rays, therefore, is only possible in models in which the annihilation proceeds via long-lived mediators, as shown in Fig. \ref{fig:dm_annihilation}. In the Sun's core, the dark matter first annihilates into a boosted long-lived mediator particle. For the models considered here, the mediator coupling to SM products is very small, which allows the mediator to be long-lived with negligible interactions in the Sun. The mediator can escape the Sun, decaying outside through observable SM channels. For a discussion of the various fields that can mediate the interaction of dark matter to photons, see Refs. \cite{2010PhRvD..81g5004B,Batell:2009yf}. For mediators that decay outside the Sun, the energy flux from dark matter annihilation is given by,

\begin{equation}
 E^{\;2}\frac{d\Phi}{dE} = \frac{\Gamma_\mathrm{ann}}{4\pi D^{2}}\;R_{i}\;E^{\;2}\frac{dN}{dE}\left(e^{-R_{\odot}/L} - e^{-D/L}\right),
 \label{eq:flux}
 \end{equation}
where $\Gamma_\mathrm{ann}$ is the rate of annihilation, $R_i$ is the branching ratio into the $i$th channel, $D$ is the distance between Sun and Earth, and $L$ is the decay length of the mediator. An important pre-requisite for an observable signal is that the mediator has a sufficiently long lifetime $\tau$ or decay length $L$, exceeding the solar radius $R_{\odot}$, so that the gamma rays are not extinguished \cite{2010PhRvD..81g5004B,2017PhRvD..95l3016L,Bell:2011sn,Arina:2017sng}. The decay length is related to the mass $m_{\chi}$ of dark matter particle, the mass $m_Y$ of the mediator, and the mediator lifetime by

\begin{equation}    
L = c\tau\frac{m_\chi}{m_{Y}}.
\end{equation}
Observations of the Sun can therefore jointly constrain the mediator lifetime and the WIMP-proton scattering cross section \cite{2017PhRvD..95l3016L}. In this work we consider the optimal case where $L \sim R_{\odot}$, such that the mediator decays just outside the Sun, producing a gamma-ray signal that would be correlated with the center of the solar disk.


\begin{figure*}[ht!]
\centering
\includegraphics[width=1.09\textwidth] {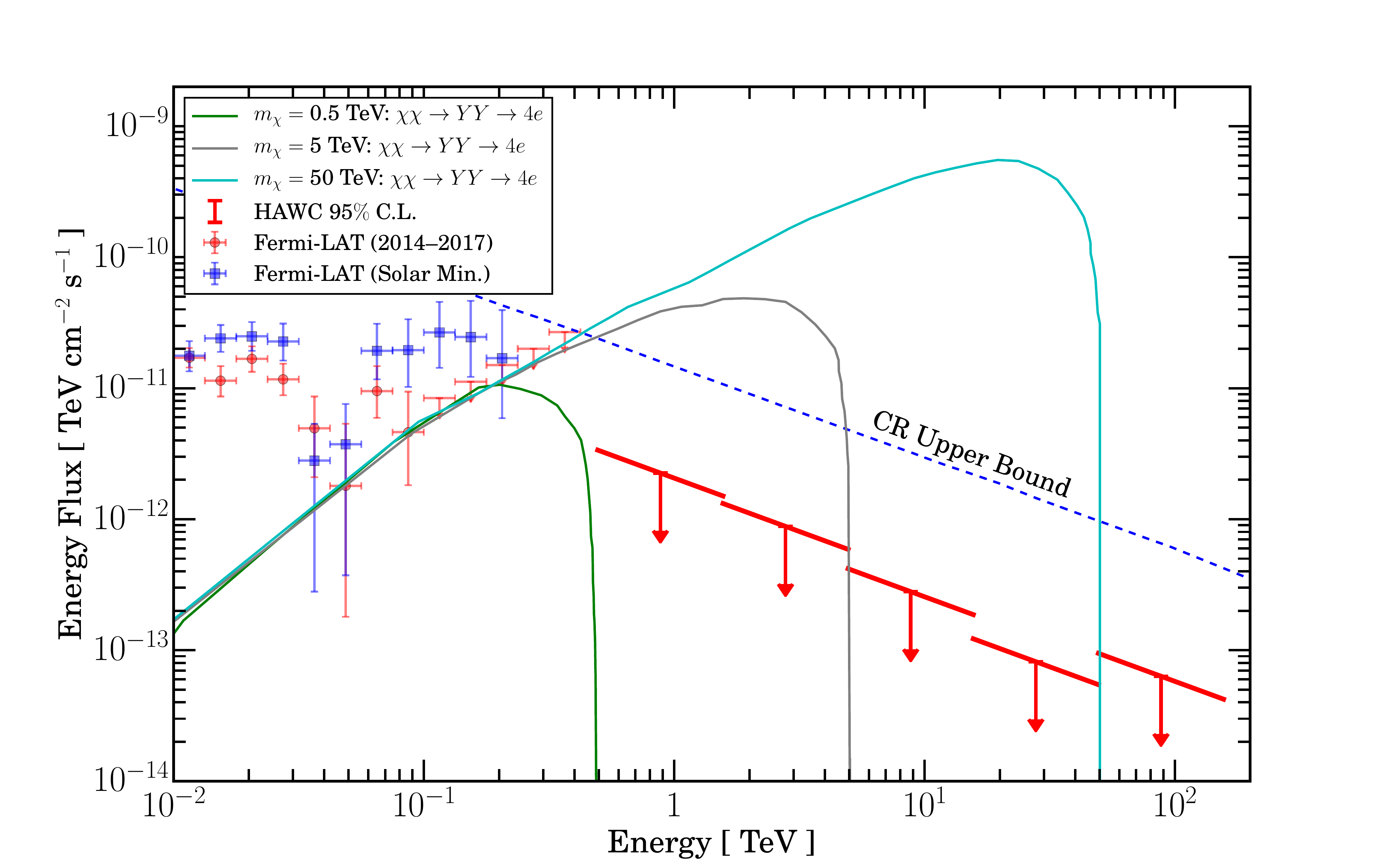}
\caption{HAWC 95\% C.L. limits on gamma-ray flux from the Sun using three years of data \cite{Astropaper}. The observed flux and the 90\% upper limits by the Fermi-LAT spanning the same period are shown in red (2014--2017) \cite{2018arXiv180406846T}. The Fermi-LAT solar min. flux was measured during the solar minimum of the solar cycle 24. The dashed line shows the theoretical maximum flux produced by hadronic interactions \cite{2018arXiv180406846T}. Also shown are a subset of predictions for various dark matter masses obtained using nominal annihilation rates allowed by the Fermi-LAT measurements, showing the power of the new HAWC limits.}
\label{fig:limits}
\end{figure*}


 
\section{\label{sec:hawc}Solar Gamma-ray Observations}
In this section, we describe the dominant astrophysical foreground for solar dark matter gamma-ray searches, and why the time window for our search is ideally situated to reduce this foreground. We also describe the GeV-TeV data sets used to set limits on the dark matter-proton spin-dependent elastic scattering cross section.

For solar dark matter searches, the sensitivity to gamma rays is accompanied by a challenge: significant foregrounds that are not well understood \cite{PhysRevD.70.083516,2011PhRvD..84c2007A,2018arXiv180406846T,2016arXiv161202420Z,Ng:2015gya,2018arXiv180305436L}. These foreground gamma rays are due to cosmic-ray interactions with solar matter and photons. The Sun has been observed in MeV-GeV gamma rays by Fermi-LAT and EGRET, leading to the identification of two distinct components \cite{Ng:2015gya,0004-637X-734-2-116,2008A&A...480..847O,2008A&A...480..847O,Orlando:2006zs,Orlando:2017iyc,2008ICRC....2..505O}: emission from the solar disk due to hadronic cosmic rays producing pions in collisions with solar gas, and a spatially extended $\sim20^{\circ}$ halo due to the inverse-Compton upscattering of solar photons by electron cosmic rays. 

A dark matter signal would be distinguishable from a cosmic-ray induced flux by its hard spectrum and a cutoff at the dark matter mass (see Fig. \ref{fig:limits}). Moreover, the flux of GeV gamma rays detected by the Fermi-LAT from the solar disk shows a distinct variability in time \cite{2018arXiv180305436L,Ng:2015gya}. The measured flux is anti-correlated with the solar activity, whereas gamma rays from dark matter annihilation should be steady in time. Thus, a search for dark matter signals from the Sun is most viable during periods of high solar activity when the foregrounds are low. As noted in Refs.~\cite{McDonald1998,2018arXiv180406846T,Ng:2015gya,2018arXiv180305436L}, the periods of relatively high solar activity correspond to a lower gamma-ray flux and a softer spectrum from the solar disk. The three-year time window considered here is expected to have a lower gamma-ray flux than during solar minimum, and hence can give stronger constraints on dark matter.

\subsection{HAWC Search for TeV Gamma Rays}

The HAWC observatory is located at an altitude of 4100 m above sea-level in the state of Puebla, Mexico. With an area of 22,000 m$^2$ and an instantaneous field-of-view of 2 sr, HAWC continuously surveys the sky for gamma rays in an energy range from $\sim$1 TeV to more than 100 TeV. HAWC consists of an array of 300 tanks; each tank contains about 200,000 liters of purified water and four photomultiplier tubes attached to its floor. Cosmic rays and gamma rays initiate particle showers in the atmosphere and produce Cherenkov light in the tanks as the particle shower passes through the water. The detection technique allows for continuous operation and gives HAWC the unique ability to observe the Sun. A complete description of the detector, data-selection and reconstruction procedure is given in Refs. \cite{2017ApJ...843...39A,Abeysekara:2013qka,Abeysekara:2017hyn,2017arXiv171000890H}.  

We analyzed data collected by HAWC between November 2014 and December 2017 \cite{Astropaper}. The HAWC period of observation falls in the second half of solar cycle 24, when the Sun is slowly becoming less active over time as it approaches the upcoming solar minimum. In a companion paper, we present the details of the measurement and the sensitivity of HAWC to TeV gamma rays from the Sun \cite{Astropaper}. We focus on gamma rays from the solar disk and, with no significant detection, our analysis rules out a gamma-ray flux above a few times $10^{-12}$ TeV$^{-1}$ cm$^{-2}$ s$^{-1}$ at near 1 TeV, approaching a sensitivity near $10\%$ of the flux from the Crab nebula. Figure \ref{fig:limits} shows constraints on the energy flux obtained by HAWC.
\subsection{Fermi-LAT Search for GeV Gamma Rays}

We also use GeV data from Fermi-LAT observations of the Sun, covering the same time period as HAWC data (Fig. \ref{fig:limits}).  The observed gamma-ray flux and upper limits from Fermi-LAT up to 400 GeV outside the solar minimum \cite{2018arXiv180406846T,2018arXiv180305436L} allow us to further constrain the annihilation rates studied in Ref.~\cite{2017PhRvD..95l3016L}.  With the updated GeV results from Fermi-LAT and TeV limits from HAWC, we are able to compute cross section limits for dark matter masses between 4 and 10$^6$ GeV. 


\begin{figure*}[ht!]
 \centering
 \makebox[\textwidth][c]{\begin{tabular}{@{}cc@{}}
 \includegraphics[width=0.55\textwidth]{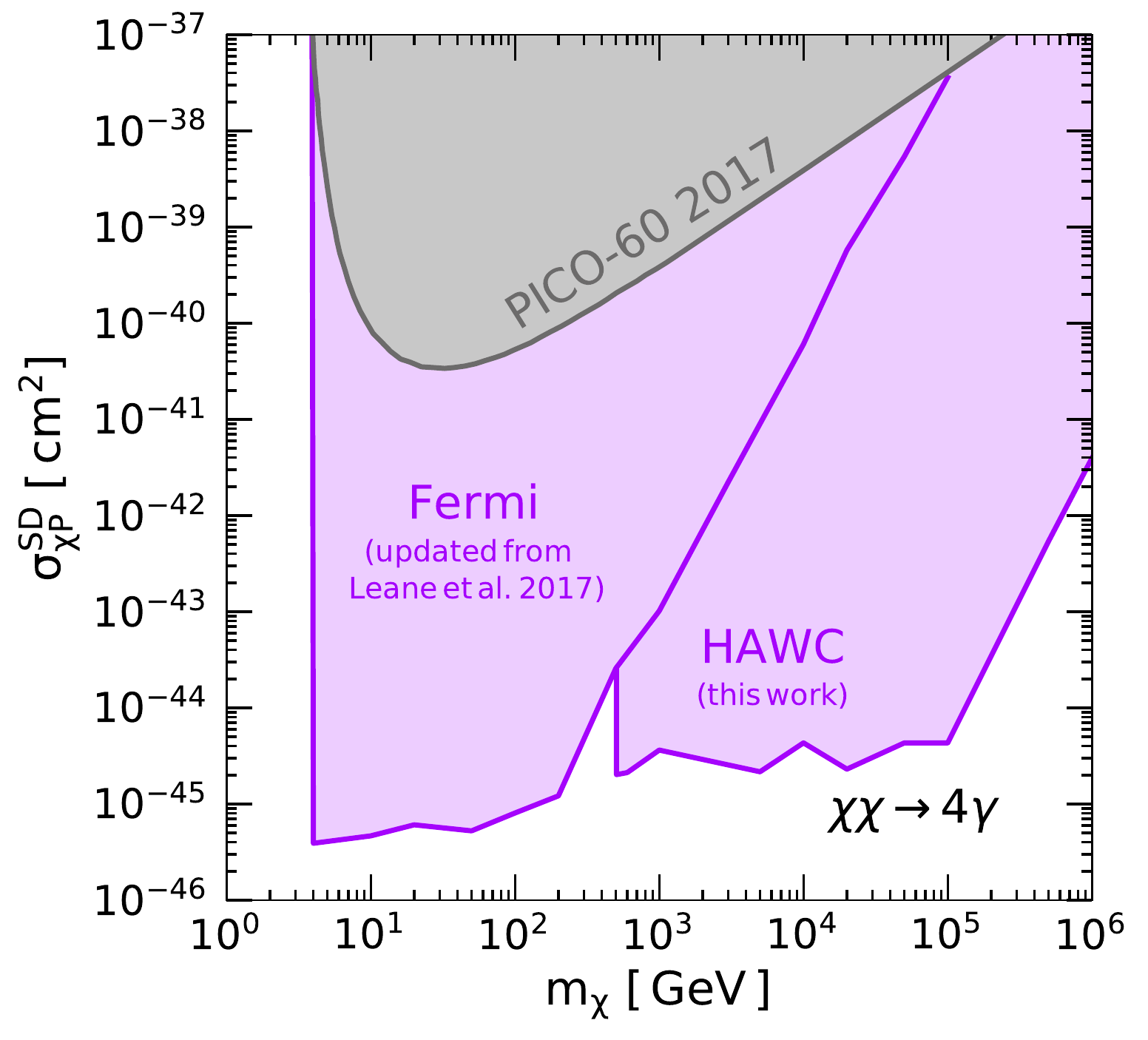} &
 \includegraphics[width=0.55\textwidth]{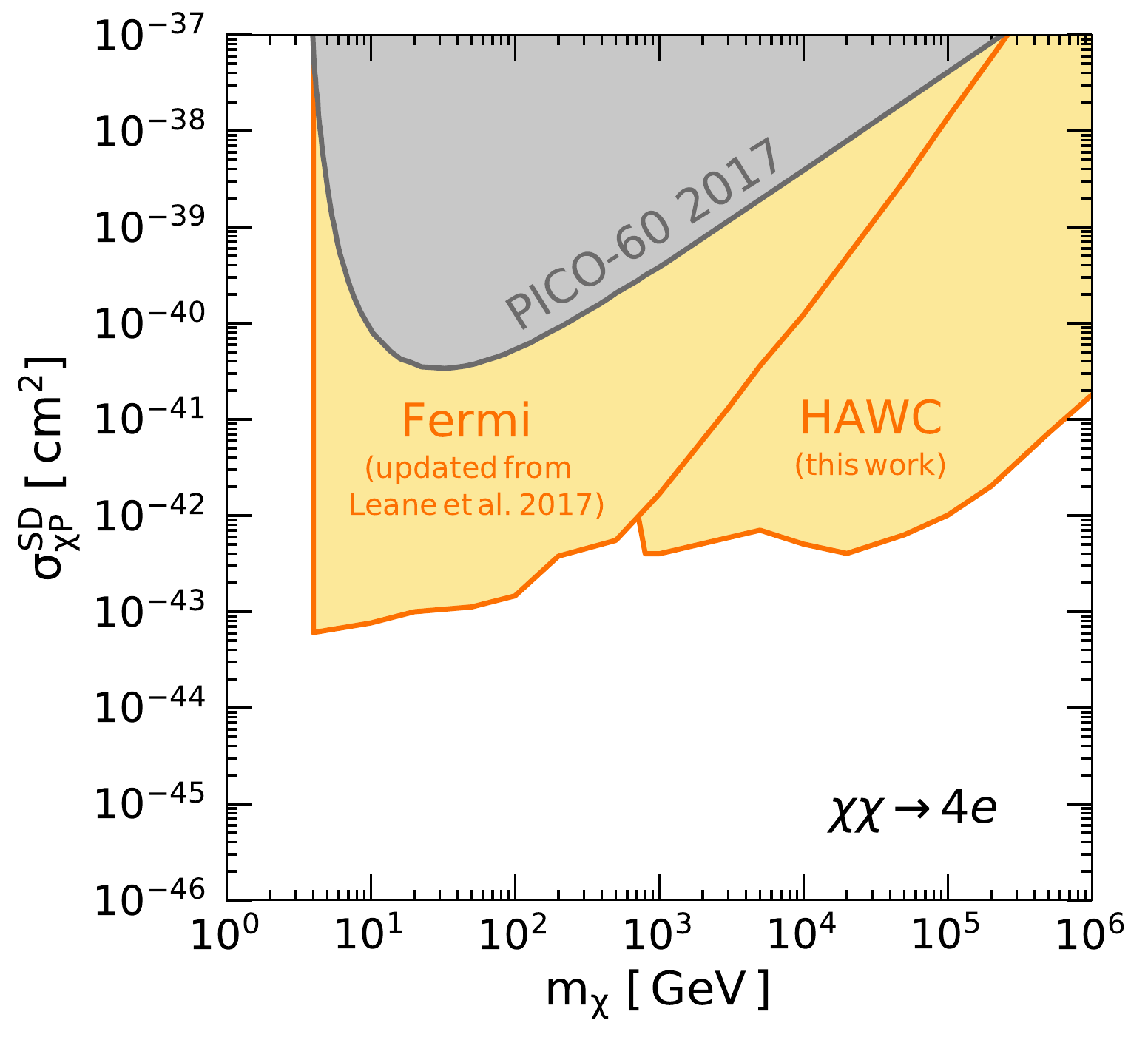} \\
 \includegraphics[width=0.55\textwidth]{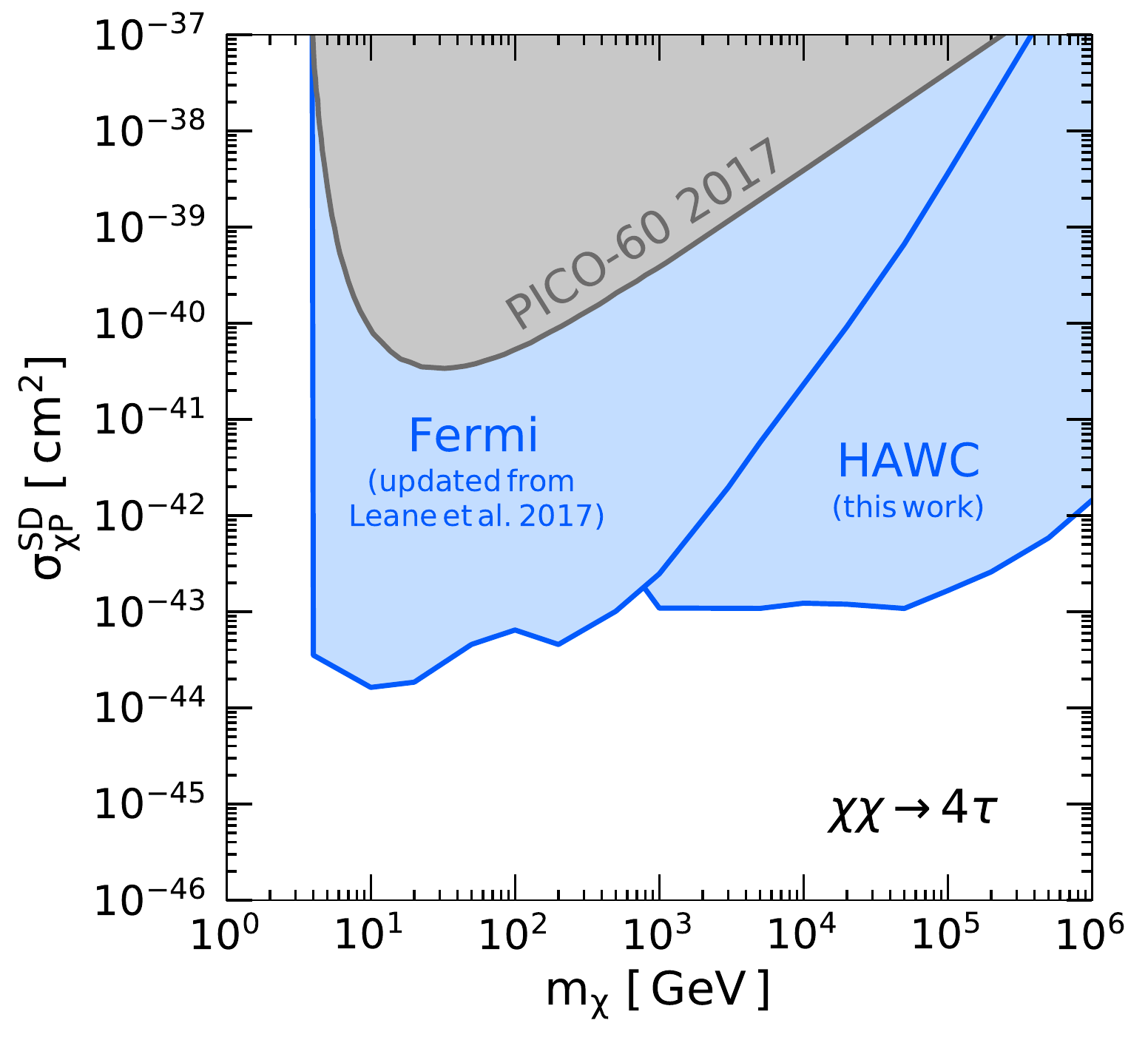}&
 \includegraphics[width=0.55\textwidth]{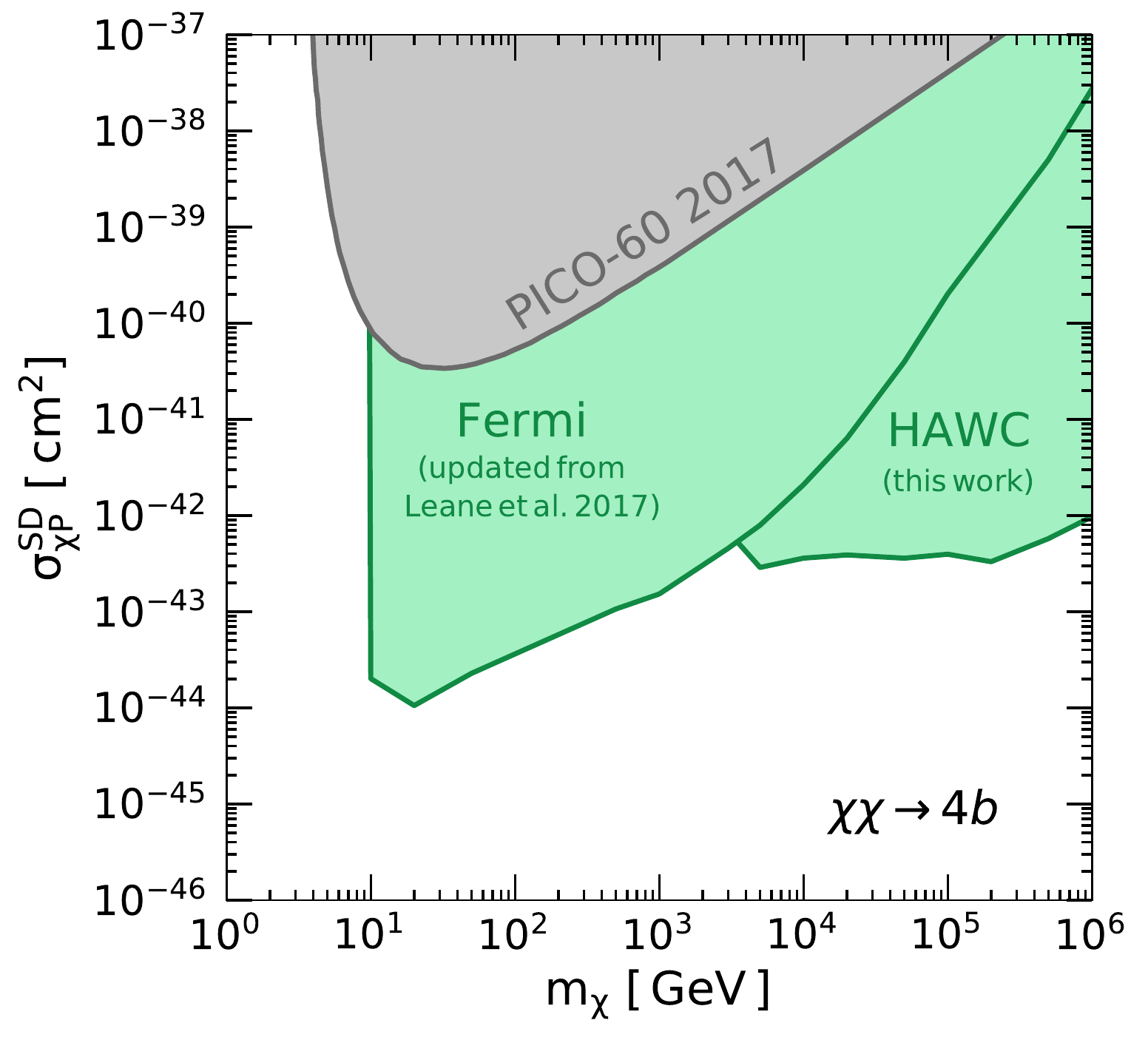}
 \end{tabular}}
 \caption{The dark matter-proton spin-dependent cross section $\sigma^{\text{SD}}$ for annihilation into pairs of $\bar{b}b$, $e^{+}e$, $\tau^+\tau^-$, and $\gamma\gamma$, assuming an optimal mediator decay length equal to the solar radius; in less favorable models, which remain to be explored, the limits would be weaker. The Fermi-LAT constraints are updated from Ref. \cite{2017PhRvD..95l3016L} using gamma-ray data from the Sun in the solar maximum (2014--2017). Also shown are the strongest direct detection constraints, obtained from PICO-60 \cite{2017PhRvL.118y1301A}.}
 \label{fig:const}
 \end{figure*}

\section{Dark Matter Constraints}
\label{constraints}
Here we present our results for dark matter scenarios with long-lived mediators, in light of the gamma-ray data described in Sec. \ref{sec:hawc}. We calculate new constraints on the spin-dependent cross section of dark matter with protons.
\subsection{\label{subsec:dmsig}Calculated Dark Matter Signals}

To set limits on the WIMP-proton scattering cross section $\sigma^{\text{SD}}$, we use Eq. (\ref{eq:flux}) and the HAWC constraints on the gamma-ray flux, for a given mediator lifetime, dark matter mass and branching ratio. We note that this is a conservative calculation based on letting the signal be 100\% of the data or limits, depending on which is available in the energy region of interest.

Assuming equilibrium has been reached, the annihilation rate $\Gamma_{\rm ann}$ of
solar dark matter is related to the capture rate as $\Gamma_{\rm ann}=\frac{1}{2}\Gamma_{\rm cap}$. We note that the standard thresholds of masses and cross sections for which equilibrium is reached~\cite{Baum:2016oow} can be substantially extended by including Sommerfeld enhancement~\cite{1931AnP...403..257S} of the dark matter annihilation rate in the Sun~\cite{Feng:2015hja}. This is naturally present in the case that the mediator is sufficiently light, which is a common property of long-lived mediators. We use an implementation of \textsc{DarkSUSY}~\cite{Gondolo:2004sc,Bringmann:2018lay} to compute the annihilation rate under this assumption, for a given dark matter scattering cross section and mass.

We generate the dark matter energy spectra using \textsc{Pythia8}~\cite{Sjostrand:2014zea}. We simulate an effective resonance with twice the dark matter energy, which decays to two mediators $YY$, which then decay to two SM final states. These SM states are either gamma rays themselves ($\chi\chi\rightarrow YY\rightarrow 4\gamma$), or produce gamma rays via radiation, or hadronic decays. We simulate the fully decayed spectra in vacuum. 
We assume $100\%$ branching fractions into individual SM final states. The limits on individual states can be rescaled individually, however for a full model with combinations of varying final states, the limits cannot just be linearly added. This is because the sum of the gamma-ray energy spectra for several final states will non-linearly change the energy in each energy bin, which non-linearly changes the overall limits.

Our approach is independent of mediator mass, provided that the mediator has sufficient boost factor $\gamma=m_\chi/m_Y$ to escape the Sun, i.e.,
\begin{equation}
 L=\gamma\beta\tau\simeq\gamma c \tau>R_\odot.
\end{equation}
There is also the possibility that the mediator is produced at an angle away from the Earth, and the gamma rays are absorbed by the Sun. Such mediator gamma rays will contribute only to the lower energy part of the dark matter energy spectrum. Our limits are set primarily by the high-energy part of the dark matter spectrum, so such effects do not affect the results.

We emphasize that we are studying the most optimistic scenario. We assume the mediator decays just outside the Sun, i.e., $L\gtrsim R_{\odot}$. More conservative scenarios can be explored by scaling the expected gamma-ray flux according to the exponential probability of signal survival discussed in Ref.~\cite{2017PhRvD..95l3016L}. The gamma-ray flux varies by only a factor of $\sim2$ across the target decay lengths between the solar surface and the Earth (see Ref.~\cite{2017PhRvD..95l3016L}). For $L> 1$ AU, the expected signal weakens exponentially, decreasing by a factor of $\sim4$ at 10 AU, and by $\sim40$ at 100 AU relative to the optimal scenario. For decay lengths less than $R_{\odot}$, the mediators do not escape the Sun, and we have no sensitivity to this scenario. Therefore, the potential gain for optimal long-lived mediators is several orders of magnitude greater than the usual short-lived mediator scenario, assuming sensitivity to neutrinos. The gamma rays cannot otherwise be probed.

As such, we assume as per the optimal scenario that the signal strength only depends on $\gamma c \tau$. 
This also means that the Sun is considered as a point source. While the angular resolution of HAWC \cite{Abeysekara:2017mjj} at high energies, is better than the $0.5^\circ$ angular diameter of the Sun, our analysis conservatively uses a larger region of interest to account for Sun-shadow effects, effectively studying it as a point source \cite{Astropaper}. So long as the mediator is highly boosted ($m_{\chi} \gg m_{Y}$, as we assume throughout) the decay products will move radially from the Sun, producing point-source emission as observed from Earth. If the mediator is not highly boosted, the size of the emission region depends on the mediator lifetime. Longer-lived, slowly moving mediators would produce more diffuse signals that would decrease HAWC's sensitivity to the solar gamma-ray signal \cite{2010PhRvD..82k5012S}.

We assume the dominant dark matter annihilation mode is two mediators that decay to SM states. This generically produces the same dark matter energy spectra in \textsc{Pythia8}, regardless of the model properties such as mass and spin, provided the mediator is sufficiently boosted. However, in some specific dark matter models, different topologies may dominate. For example, if the long-lived mediator is a pseudoscalar, the two-mediator annihilation mode is $p$-wave suppressed, and instead the $s$-wave $\chi\chi\rightarrow YYY$ may dominate~\cite{Abdullah:2014lla, Rajaraman:2015xka,Bell:2017irk}, leading to a different spectral energy distribution. In such scenarios, the upper limits will be different from what we found for the optimal case. Moreover, while high mediator boosts are achieved for all dark matter masses in this work with electrons and gamma rays as final states, for  taus and b-quarks, large dark matter masses are required for a sufficiently boosted mediator. However, as the cutoff mass is semi-arbitrary, we show results for all masses where the direct decays are kinematically allowed, and note the resulting weakening of limits for heavier final states if the mediator is not highly boosted \cite{2017PhRvD..95l3016L}. 

The limits would also be weaker if the assumed decay length was much shorter or longer than the optimal case. For $L<R_{\odot}$, there would be attenuation inside the Sun, with essentially any depth under the surface extinguishing the gamma rays, so that the reduced signal would just be the portion decaying outside the Sun. For $L\gg R_{\odot}$, the sensitivity declines with the flux loss as per Eq. \ref{eq:flux}.

Our goal is not the detailed exploration of specific models. Instead, we demonstrate the power of the first strong constraints on solar TeV gamma rays as a probe of spin-dependent dark matter-proton scattering. With the optimal scenario considered here, we go several orders of magnitude below what is presently constrained by direct searches. 


\begin{figure*}[t!]
\centering
\includegraphics[width=0.99\textwidth] {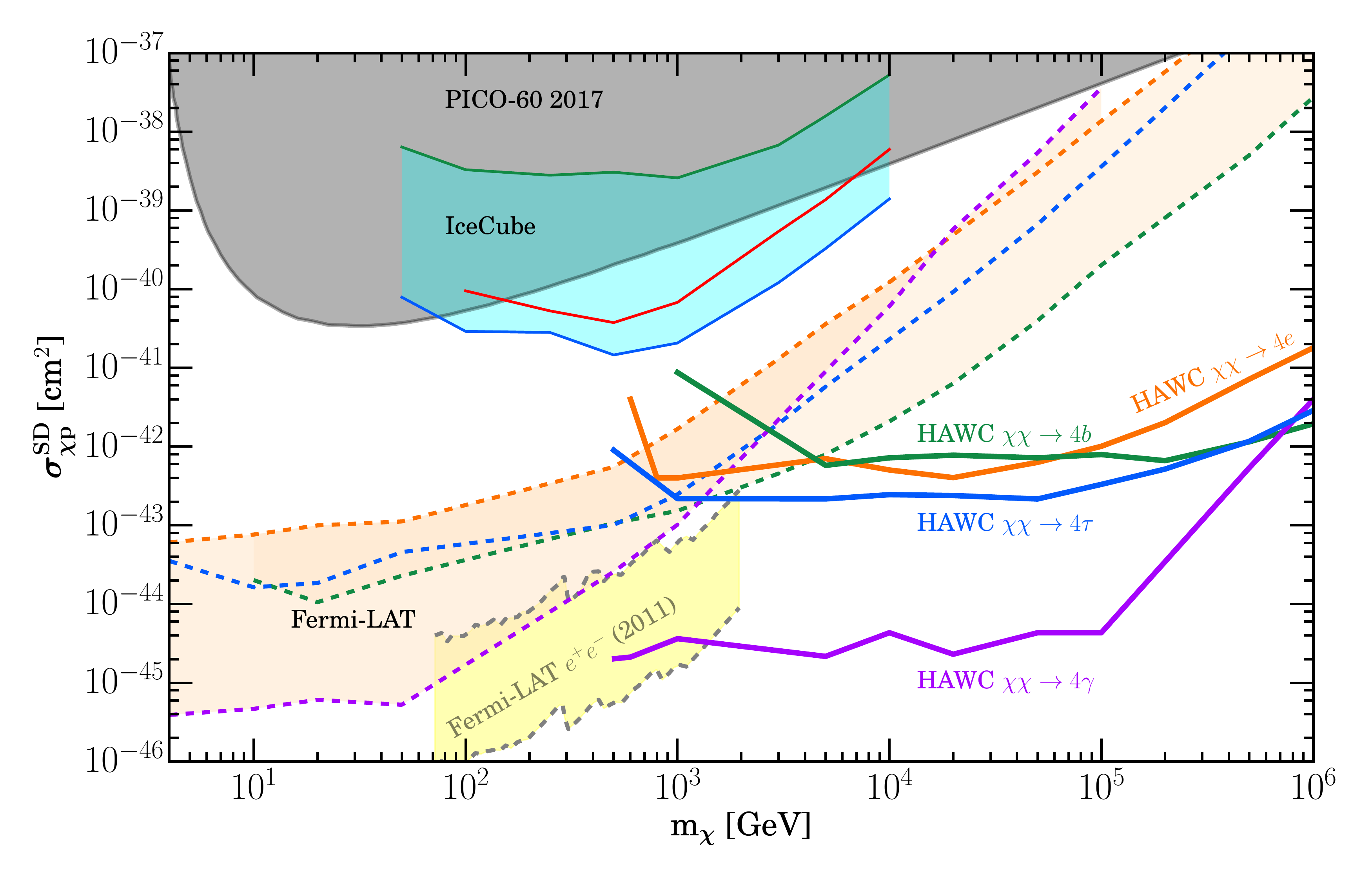}
\caption{Upper limits on the spin-dependent dark matter-proton scattering cross section $\sigma^{\text{SD}}$, from various gamma-ray and neutrino experiments. The thick solid lines show the limits obtained from HAWC in this work for $\bar{b}b$, $e^{+}e$, $\tau^+\tau^-$ and $\gamma\gamma$ channels. The dashed lines following the same channel-wise color scheme show the limits from Fermi-LAT for $\bar{b}b$, $e^{+}e$, $\tau^+\tau^-$ and $\gamma\gamma$, updated from Ref. \cite{2017PhRvD..95l3016L}. The HAWC and the Fermi-LAT limits in this work are for scenarios where the mediator has a decay length $L = R_{\odot}$, and can be scaled for more conservative cases as discussed in Sec. \ref{subsec:dmsig}. The 2011 Fermi-LAT results are for mediators decaying into electrons, with decay lengths between 0.1 and 5 AU \cite{Ajello:2011dq}. The thin solid lines show the results from IceCube for the $\bar{b}b$, $W^{+}W^{-}$, $\tau^+\tau^-$channels \cite{2017EPJC...77..146A}, and are for scenarios with short-lived mediators (results for long-lived mediators would be better \cite{Bell:2011sn,2017PhRvD..95l3016L}, but IceCube results are not yet available). The grey region indicates the parameter space excluded by PICO-60 \cite{2017PhRvL.118y1301A}.} 
\label{fig:all_dm}
\end{figure*}


\subsection{Limits on Spin-Dependent Dark Matter Scattering}

The limits we present on spin-dependent dark matter scattering require the presence of a sufficiently long-lived dark mediator, for the produced gamma rays to escape the solar surface. Dark matter captured in the Sun predominantly annihilates to two on-shell mediators (annihilation to only one mediator is phase-space suppressed). Each of these mediators decays to two SM final states, leading to a total of four SM final states. We chose four representative cases for SM final states, due to their varying spectral energy distributions: $4\gamma$ (box spectrum, hardest), $4\tau$ (hard spectrum), $4e$ and $4b$ (softer spectra). 

Figure \ref{fig:const} shows the constraints on the spin-dependent scattering cross section obtained here for gamma rays from mediator decays to the $\gamma\gamma$, $e^{+}e^-$,  $\tau^+\tau^-$, and $\overline{b}b$ channels. For any given channel, the constraints provided by HAWC are at least three to four orders of magnitude stronger than the strongest limits provided by direct-detection methods, for dark matter masses above 1 TeV (at high masses, the gain is much more than that). Compared to a previous study of Milagro sensitivity for constraining long-lived mediator scenarios with TeV solar gamma rays, the HAWC constraints are about three orders of magnitude stronger \cite{2010PhRvD..81g5004B}. 

Previously, Fermi-LAT has also searched for long-lived mediators decaying into electrons outside the Sun, and the resulting constraints for dark matter masses between 70 and 2000 GeV are stronger than the limits from solar gamma rays \cite{Ajello:2011dq}. While the limits in Ref. \cite{Ajello:2011dq} are set by a non-observation of electrons from the Sun, our analysis uses the observed gamma rays from Fermi-LAT, and requires 100$\%$ of the flux to contribute to the expected dark matter signal, which yields a less constraining but conservative result. Moreover, in this work we only consider the gamma rays produced from dark matter annihilation. For the electron final state, the gamma rays are subdominant because they are produced through bremsstrahlung. In principle, the HAWC measurement may also include a flux of electrons from the Sun which cannot be differentiated from the gamma rays. Adding the electron contribution would only improve our results, and could make the $\chi\chi \rightarrow 4e$ channel as strong as that of $\chi\chi \rightarrow 4\gamma$. Doing so would require an estimate of the electron deflection angle from the Sun to the Earth, which is beyond the scope of the current work. 

For a given channel, the constraints become weaker at higher dark matter mass. This is because the rate of capture and hence annihilation decreases for $m_{\chi} \gtrsim100~\mathrm{GeV}$ \cite{2014JCAP...05..049C}. The differences between the limits for different channels at the highest energies, depend on their spectral energy distributions relative to the HAWC sensitivity.
Note that for a generic WIMP, unitarity and bound state effects become important around 100 TeV~\cite{PhysRevLett.64.615, Blum:2014dca,vonHarling:2014kha, Harigaya:2016nlg,  Bramante:2017obj, Baldes:2017gzw}. For model-dependent choices, such constraints and effects should be taken into account for the heaviest masses we consider.

We show constraints on the spin-dependent scattering cross section from Fermi-LAT, updated from Ref.~\cite{2017PhRvD..95l3016L} to match the time period with HAWC (away from the solar minimum). Fermi-LAT's GeV measurements are complementary to HAWC's TeV measurements, providing the strongest constraints for $m_{\chi} < 1$ TeV, and becoming less sensitive at higher dark matter masses. Together, Fermi-LAT and HAWC measurements provide powerful bounds for dark matter masses between 4 GeV and 10$^6$ GeV. 
 
Figure \ref{fig:all_dm} shows the constraints on the spin-dependent scattering cross section obtained with HAWC and Fermi-LAT compared to other sensitive direct and indirect detection experiments. Among the direct-detection experiments, the most stringent constraints are provided by PICO 60 - $C_3F_8$ \cite{2017PhRvL.118y1301A}. Unlike the Xe-based detectors that have only a fraction of isotopes sensitive to spin-dependent scattering, PICO uses fluorine as the target nucleus which, due to its odd number of nucleons, is more sensitive to spin-dependent scattering. Also shown are the complementary neutrino channel limits from IceCube for general WIMP scenarios, where the dark matter neutrino signal could originate from the Sun's core without an intermediate long-lived mediator \cite{2017EPJC...77..146A}. With long-lived mediators, there is less neutrino attenuation and the resulting constraints from neutrino telescopes can be much stronger above 1 TeV \cite{Bell:2011sn,2017PhRvD..95l3016L,Adrian-Martinez:2016ujo}. 

For TeV-scale dark matter with long-lived mediators, both gamma-ray and neutrino searches are prone to foregrounds due to the astrophysical emission from the Sun. This astrophysical emission includes solar atmospheric neutrinos \cite{2017PhRvD..96j3006N,2017JCAP...07..024A,Edsjo:2017kjk,2018APh....97...63M} and gamma rays from cosmic-ray interactions (Sec. \ref{sec:hawc}). Once the sensitivity is good enough to detect the astrophysical flux, further improvements become more difficult, i.e., a (soft) sensitivity floor has been reached. For neutrinos, the astrophysical foreground may soon be detectable \cite{icecubesolar} and be indistinguishable from a dark matter signal due to the limited energy resolution of neutrino telescopes at the relevant energies \cite{0954-3899-19-9-019,Ingelman:1996mj,2017PhRvD..96j3006N,2017JCAP...07..024A,2011JCAP...09..029R,Edsjo:2017kjk,Danninger:2014xza}. For gamma rays, the astrophysical foreground at TeV energies is still unknown \cite{Astropaper}. The minimum flux of gamma rays from cosmic-ray interactions with the solar limb \cite{2016arXiv161202420Z,Gao:2017bfv} is three orders of magnitude below the upper bound shown in Fig. \ref{fig:limits}. There is significant room for improvement in sensitivity to gamma rays from the Sun before the floor is reached. Even then, gamma rays from cosmic-ray interactions could be distinguished from dark matter based on their unique spectrum and time-variability.

Collider searches for missing-momentum attributed to dark matter production can also be translated to limits on the dark matter-proton scattering cross section, by making some model dependent choices~\cite{Abercrombie:2015wmb, Albert:2017onk}. Both ATLAS \cite{Aaboud:2017phn} and CMS~\cite{Sirunyan:2017hci} have recast their limits in the case where dark matter-quark interactions are mediated by an axial-vector, with coupling 0.25 to quarks and 1 to dark matter, obtaining limits on the spin-dependent scattering cross section up to about $10^{-42}$ cm$^2$ for dark matter masses less than around a few hundred GeV. While these limits hold only for this specific model and parameter choices, in general, collider limits are complementary to those we obtain in this work from HAWC, which is most sensitive to higher dark matter masses.

\section{\label{sec:conc}Conclusions}
Dark matter capture and annihilation in the Sun provides a probe of the dark matter-proton scattering cross section. In the presence of sufficiently long-lived dark mediators, the gamma rays produced can also escape the Sun, providing new ways to detect dark matter.

We have demonstrated that gamma-ray measurements of the Sun are exceptionally sensitive to the dark matter scattering cross section. Using new data from HAWC's search for gamma rays from the solar disk, we place the strongest existing constraints on TeV dark matter and its spin-dependent scattering rate, assuming an optimal long-lived mediator lifetime. We also update limits from Fermi-LAT with data during the same period of observation, placing the strongest existing constraints on spin-dependent scattering of GeV dark matter. Together, Fermi-LAT and HAWC place severe bounds on spin-dependent dark matter scattering, for dark matter masses between 4 GeV and 10$^6$ GeV. Under optimal model assumptions, this reaches down to scattering cross sections of about $10^{-45}$ cm$^2$, outperforming leading direct detection experiments by many orders of magnitude, especially for large dark matter masses.

Long-lived mediators are present naturally in many new physics models, especially if a dark mediator has small couplings or a low mass. Our first strong constraints on solar TeV gamma rays provide a new, powerful way to probe theoretical models of long-lived mediators, along with the nature of dark matter. The constraints are the strongest across a wide range of dark matter masses, and will be important for future studies of new physics. These new bounds invite detailed exploration of which model-specific realizations of the long-lived mediator scenario are not eliminated.

\vspace{-0.5cm}
\section*{Acknowledgments}
\vspace{-0.3cm}
We acknowledge the support from: the US National Science Foundation (NSF) the US Department of Energy Office of High-Energy Physics;  the Laboratory Directed Research and Development (LDRD) program of Los Alamos National Laboratory; Consejo Nacional de Ciencia y Tecnolog\'{\i}a (CONACyT), M{\'e}xico (grants 271051, 232656, 260378, 179588, 239762, 254964, 271737, 258865, 243290, 132197, 281653 (C{\'a}tedras 873,1563, 341), Laboratorio Nacional HAWC de rayos gamma; L'OREAL Fellowship for Women in Science 2014; Red HAWC, M{\'e}xico; DGAPA-UNAM (grants IG100317, IN111315, IN111716-3, IA102715, 109916, IA102917, IN112218); VIEP-BUAP; PIFI 2012, 2013, PROFOCIE 2014, 2015; the University of Wisconsin Alumni Research Foundation; the Institute of Geophysics, Planetary Physics, and Signatures at Los Alamos National Laboratory; Polish Science Centre grant DEC-2014/13/B/ST9/945, DEC-2017/27/B/ST9/02272; Coordinaci{\'o}n de la Investigaci{\'o}n Cient\'{\i}fica de la Universidad Michoacana; Royal Society - Newton Advanced Fellowship 180385. Thanks to Scott Delay, Luciano D\'{\i}az and Eduardo Murrieta for technical support.

JFB is supported by (and BZ is partially supported by) NSF grant PHY-1714479. RKL is supported by the Office of High Energy Physics of the U.S. Department of Energy under Grant No.\;DE-SC00012567 and DE-SC0013999. KCYN is supported by Croucher Fellowship and Benoziyo Fellowship.
TL, AHGP and BZ are supported in part by NASA Grant No.\;80NSSC17K0754.

\bibliography{bib}{}
\end{document}

%% file: authorlist.tex
\author{A.~Albert}\affiliation{Physics Division, Los Alamos National Laboratory, Los Alamos, NM, USA }
\author{R.~Alfaro}\affiliation{Instituto de F\'{i}sica, Universidad Nacional Autónoma de México, Ciudad de Mexico, Mexico }
\author{C.~Alvarez}\affiliation{Universidad Autónoma de Chiapas, Tuxtla Gutiérrez, Chiapas, México}
\author{R.~Arceo}\affiliation{Universidad Autónoma de Chiapas, Tuxtla Gutiérrez, Chiapas, México}
\author{J.C.~Arteaga-Velázquez}\affiliation{Universidad Michoacana de San Nicolás de Hidalgo, Morelia, Mexico }
\author{D.~Avila Rojas}\affiliation{Instituto de F\'{i}sica, Universidad Nacional Autónoma de México, Ciudad de Mexico, Mexico }
\author{H.A.~Ayala Solares}\affiliation{Department of Physics, Pennsylvania State University, University Park, PA, USA }
\author{E.~Belmont-Moreno}\affiliation{Instituto de F\'{i}sica, Universidad Nacional Autónoma de México, Ciudad de Mexico, Mexico }
\author{S.Y.~BenZvi}\affiliation{Department of Physics \& Astronomy, University of Rochester, Rochester, NY , USA }
\author{C.~Brisbois}\affiliation{Department of Physics, Michigan Technological University, Houghton, MI, USA }
\author{K.S.~Caballero-Mora}\affiliation{Universidad Autónoma de Chiapas, Tuxtla Gutiérrez, Chiapas, México}
\author{T.~Capistrán}\affiliation{Instituto Nacional de Astrof\'{i}sica, Óptica y Electrónica, Puebla, Mexico }
\author{A.~Carramiñana}\affiliation{Instituto Nacional de Astrof\'{i}sica, Óptica y Electrónica, Puebla, Mexico }
\author{S.~Casanova}\affiliation{Institute of Nuclear Physics Polish Academy of Sciences, PL-31342 IFJ-PAN, Krakow, Poland }
\author{M.~Castillo}\affiliation{Universidad Michoacana de San Nicolás de Hidalgo, Morelia, Mexico }
\author{J.~Cotzomi}\affiliation{Facultad de Ciencias F\'{i}sico Matemáticas, Benemérita Universidad Autónoma de Puebla, Puebla, Mexico }
\author{S.~Coutiño de León}\affiliation{Instituto Nacional de Astrof\'{i}sica, Óptica y Electrónica, Puebla, Mexico }
\author{C.~De León}\affiliation{Facultad de Ciencias F\'{i}sico Matemáticas, Benemérita Universidad Autónoma de Puebla, Puebla, Mexico }
\author{E.~De la Fuente}\affiliation{Departamento de F\'{i}sica, Centro Universitario de Ciencias Exactase Ingenierias, Universidad de Guadalajara, Guadalajara, Mexico }
\author{S.~Dichiara}\affiliation{Instituto de Astronom\'{i}a, Universidad Nacional Autónoma de México, Ciudad de Mexico, Mexico }
\author{B.L.~Dingus}\affiliation{Physics Division, Los Alamos National Laboratory, Los Alamos, NM, USA }
\author{M.A.~DuVernois}\affiliation{Department of Physics, University of Wisconsin-Madison, Madison, WI, USA }
\author{J.C.~Díaz-Vélez}\affiliation{Departamento de F\'{i}sica, Centro Universitario de Ciencias Exactase Ingenierias, Universidad de Guadalajara, Guadalajara, Mexico }
\author{K.~Engel}\affiliation{Department of Physics, University of Maryland, College Park, MD, USA }
\author{O.~Enríquez-Rivera}\affiliation{Instituto de Geof\'{i}sica, Universidad Nacional Autónoma de México, Ciudad de Mexico, Mexico }
\author{C.~Espinoza}\affiliation{Instituto de F\'{i}sica, Universidad Nacional Autónoma de México, Ciudad de Mexico, Mexico }
\author{H.~Fleischhack}\affiliation{Department of Physics, Michigan Technological University, Houghton, MI, USA }
\author{N.~Fraija}\affiliation{Instituto de Astronom\'{i}a, Universidad Nacional Autónoma de México, Ciudad de Mexico, Mexico }
\author{J.A.~García-González}\affiliation{Instituto de F\'{i}sica, Universidad Nacional Autónoma de México, Ciudad de Mexico, Mexico }
\author{F.~Garfias}\affiliation{Instituto de Astronom\'{i}a, Universidad Nacional Autónoma de México, Ciudad de Mexico, Mexico }
\author{M.M.~González}\affiliation{Instituto de Astronom\'{i}a, Universidad Nacional Autónoma de México, Ciudad de Mexico, Mexico }
\author{J.A.~Goodman}\affiliation{Department of Physics, University of Maryland, College Park, MD, USA }
\author{Z.~Hampel-Arias}\affiliation{Department of Physics, University of Wisconsin-Madison, Madison, WI, USA }
\author{J.P.~Harding}\affiliation{Physics Division, Los Alamos National Laboratory, Los Alamos, NM, USA }
\author{S.~Hernandez}\affiliation{Instituto de F\'{i}sica, Universidad Nacional Autónoma de México, Ciudad de Mexico, Mexico }
\author{B.~Hona}\affiliation{Department of Physics, Michigan Technological University, Houghton, MI, USA }
\author{F.~Hueyotl-Zahuantitla}\affiliation{Universidad Autónoma de Chiapas, Tuxtla Gutiérrez, Chiapas, México}
\author{P.~Hüntemeyer}\affiliation{Department of Physics, Michigan Technological University, Houghton, MI, USA }
\author{A.~Iriarte}\affiliation{Instituto de Astronom\'{i}a, Universidad Nacional Autónoma de México, Ciudad de Mexico, Mexico }
\author{A.~Jardin-Blicq}\affiliation{Max-Planck Institute for Nuclear Physics, 69117 Heidelberg, Germany}
\author{V.~Joshi}\affiliation{Max-Planck Institute for Nuclear Physics, 69117 Heidelberg, Germany}
\author{S.~Kaufmann}\affiliation{Universidad Autónoma de Chiapas, Tuxtla Gutiérrez, Chiapas, México}
\author{H.~León Vargas}\affiliation{Instituto de F\'{i}sica, Universidad Nacional Autónoma de México, Ciudad de Mexico, Mexico }
\author{G.~Luis-Raya}\affiliation{Universidad Politecnica de Pachuca, Pachuca, Hgo, Mexico }
\author{J.~Lundeen}\affiliation{Department of Physics and Astronomy, Michigan State University, East Lansing, MI, USA }
\author{R.~López-Coto}\affiliation{INFN and Universita di Padova, via Marzolo 8, I-35131,Padova,Italy}
\author{K.~Malone}\affiliation{Department of Physics, Pennsylvania State University, University Park, PA, USA }
\author{S.S.~Marinelli}\affiliation{Department of Physics and Astronomy, Michigan State University, East Lansing, MI, USA }
\author{O.~Martinez}\affiliation{Facultad de Ciencias F\'{i}sico Matemáticas, Benemérita Universidad Autónoma de Puebla, Puebla, Mexico }
\author{I.~Martinez-Castellanos}\affiliation{Department of Physics, University of Maryland, College Park, MD, USA }
\author{J.~Martínez-Castro}\affiliation{Centro de Investigaci\'on en Computaci\'on, Instituto Polit\'ecnico Nacional, M\'exico City, M\'exico.}
\author{P.~Miranda-Romagnoli}\affiliation{Universidad Autónoma del Estado de Hidalgo, Pachuca, Mexico }
\author{E.~Moreno}\affiliation{Facultad de Ciencias F\'{i}sico Matemáticas, Benemérita Universidad Autónoma de Puebla, Puebla, Mexico }
\author{M.~Mostafá}\affiliation{Department of Physics, Pennsylvania State University, University Park, PA, USA }
\author{A.~Nayerhoda}\affiliation{Institute of Nuclear Physics Polish Academy of Sciences, PL-31342 IFJ-PAN, Krakow, Poland }
\author{L.~Nellen}\affiliation{Instituto de Ciencias Nucleares, Universidad Nacional Autónoma de Mexico, Ciudad de Mexico, Mexico }
\author{M.~Newbold}\affiliation{Department of Physics and Astronomy, University of Utah, Salt Lake City, UT, USA }
\author{M.U.~Nisa}\email{Corresponding author \\Email: mnisa@ur.rochester.edu}\affiliation{Department of Physics \& Astronomy, University of Rochester, Rochester, NY , USA }
\author{R.~Noriega-Papaqui}\affiliation{Universidad Autónoma del Estado de Hidalgo, Pachuca, Mexico }
\author{J.~Pretz}\affiliation{Department of Physics, Pennsylvania State University, University Park, PA, USA }
\author{E.G.~Pérez-Pérez}\affiliation{Universidad Politecnica de Pachuca, Pachuca, Hgo, Mexico }
\author{Z.~Ren}\affiliation{Dept of Physics and Astronomy, University of New Mexico, Albuquerque, NM, USA }
\author{C.D.~Rho}\affiliation{Department of Physics \& Astronomy, University of Rochester, Rochester, NY , USA }
\author{C.~Rivière}\affiliation{Department of Physics, University of Maryland, College Park, MD, USA }
\author{D.~Rosa-González}\affiliation{Instituto Nacional de Astrof\'{i}sica, Óptica y Electrónica, Puebla, Mexico }
\author{M.~Rosenberg}\affiliation{Department of Physics, Pennsylvania State University, University Park, PA, USA }
\author{E.~Ruiz-Velasco}\affiliation{Max-Planck Institute for Nuclear Physics, 69117 Heidelberg, Germany}
\author{H.~Salazar}\affiliation{Facultad de Ciencias F\'{i}sico Matemáticas, Benemérita Universidad Autónoma de Puebla, Puebla, Mexico }
\author{F.~Salesa Greus}\affiliation{Institute of Nuclear Physics Polish Academy of Sciences, PL-31342 IFJ-PAN, Krakow, Poland }
\author{A.~Sandoval}\affiliation{Instituto de F\'{i}sica, Universidad Nacional Autónoma de México, Ciudad de Mexico, Mexico }
\author{M.~Schneider}\affiliation{Department of Physics, University of Maryland, College Park, MD, USA }
\author{M.~Seglar Arroyo}\affiliation{Department of Physics, Pennsylvania State University, University Park, PA, USA }
\author{G.~Sinnis}\affiliation{Physics Division, Los Alamos National Laboratory, Los Alamos, NM, USA }
\author{A.J.~Smith}\affiliation{Department of Physics, University of Maryland, College Park, MD, USA }
\author{R.W.~Springer}\affiliation{Department of Physics and Astronomy, University of Utah, Salt Lake City, UT, USA }
\author{P.~Surajbali}\affiliation{Max-Planck Institute for Nuclear Physics, 69117 Heidelberg, Germany}
\author{I.~Taboada}\affiliation{School of Physics and Center for Relativistic Astrophysics - Georgia Institute of Technology, Atlanta, GA, USA 30332 }
\author{O.~Tibolla}\affiliation{Universidad Autónoma de Chiapas, Tuxtla Gutiérrez, Chiapas, México}
\author{K.~Tollefson}\affiliation{Department of Physics and Astronomy, Michigan State University, East Lansing, MI, USA }
\author{I.~Torres}\affiliation{Instituto Nacional de Astrof\'{i}sica, Óptica y Electrónica, Puebla, Mexico }
\author{L.~Villaseñor}\affiliation{Facultad de Ciencias F\'{i}sico Matemáticas, Benemérita Universidad Autónoma de Puebla, Puebla, Mexico }
\author{T.~Weisgarber}\affiliation{Department of Physics, University of Wisconsin-Madison, Madison, WI, USA }
\author{S.~Westerhoff}\affiliation{Department of Physics, University of Wisconsin-Madison, Madison, WI, USA }
\author{I.G.~Wisher}\affiliation{Department of Physics, University of Wisconsin-Madison, Madison, WI, USA }
\author{J.~Wood}\affiliation{Department of Physics, University of Wisconsin-Madison, Madison, WI, USA }
\author{T.~Yapici}\affiliation{Department of Physics \& Astronomy, University of Rochester, Rochester, NY , USA }
\author{A.~Zepeda}\affiliation{Physics Department, Centro de Investigacion y de Estudios Avanzados del IPN, Mexico City, DF, Mexico }
\author{H.~Zhou}\affiliation{Physics Division, Los Alamos National Laboratory, Los Alamos, NM, USA }
\author{J.D.~Álvarez}\affiliation{Universidad Michoacana de San Nicolás de Hidalgo, Morelia, Mexico }

%% file: guest_author.tex
\author{J. F. Beacom}
\affiliation{Center for Cosmology and AstroParticle Physics (CCAPP), Ohio State University, Columbus, Ohio 43210, USA}
\affiliation{Department of Physics, Ohio State University, Columbus, Ohio 43210, USA}
\affiliation{Department of Astronomy, Ohio State University, Columbus, Ohio 43210, USA}

\author{R. K.\ Leane}
\affiliation{Center for Theoretical Physics, Massachusetts Institute of Technology, Cambridge, MA 02139, USA}

\author{T. Linden}
\affiliation{Center for Cosmology and AstroParticle Physics (CCAPP), Ohio State University, Columbus, Ohio 43210, USA}

\author{K. C. Y. Ng}
\affiliation{Department of Particle Physics and Astrophysics, Weizmann Institute of Science, Rehovot 76100, Israel}

\author{A. H. G. Peter}
\affiliation{Center for Cosmology and AstroParticle Physics (CCAPP), Ohio State University, Columbus, Ohio 43210, USA}
\affiliation{Department of Physics, Ohio State University, Columbus, Ohio 43210, USA}
\affiliation{Department of Astronomy, Ohio State University, Columbus, Ohio 43210, USA}

\author{B. Zhou \vspace{0.5cm}}
\affiliation{Center for Cosmology and AstroParticle Physics (CCAPP), Ohio State University, Columbus, Ohio 43210, USA}
\affiliation{Department of Physics, Ohio State University, Columbus, Ohio 43210, USA}

%% file: mediator.tex
\begin{tikzpicture}{remember picture, overlay}
  %
  %
	\foreach\i in {0.5,0.51,...,1} {
		\fill[opacity=\i*0.15,orange] (0,0) circle ({6-6*\i});
	}
  %
  %
  \draw (0,0.3) node {$\chi\chi\rightarrow$ mediators};
  \draw[fill=black] (0,0) circle (0.5mm);
  \draw[dashed] (0,0) -- (3,-1.3);
  \draw[fill=black] (3,-1.3) circle (0.5mm);
  \draw[->] (3,-1.3) -- (4,-1) node [right] {$\nu$};
  \draw[->] (3,-1.3) -- (4.2,-1.7) node [right] {$\gamma$};
  \draw[->] (3,-1.3) -- (4,-2.3) node [right] {$e^{\pm}$, $\mu^{\pm}$, $\ldots$};
\end{tikzpicture}

%% file: dm_v2submit.bbl
\begin{thebibliography}{115}%
\makeatletter
\providecommand \@ifxundefined [1]{%
 \@ifx{#1\undefined}
}%
\providecommand \@ifnum [1]{%
 \ifnum #1\expandafter \@firstoftwo
 \else \expandafter \@secondoftwo
 \fi
}%
\providecommand \@ifx [1]{%
 \ifx #1\expandafter \@firstoftwo
 \else \expandafter \@secondoftwo
 \fi
}%
\providecommand \natexlab [1]{#1}%
\providecommand \enquote  [1]{``#1''}%
\providecommand \bibnamefont  [1]{#1}%
\providecommand \bibfnamefont [1]{#1}%
\providecommand \citenamefont [1]{#1}%
\providecommand \href@noop [0]{\@secondoftwo}%
\providecommand \href [0]{\begingroup \@sanitize@url \@href}%
\providecommand \@href[1]{\@@startlink{#1}\@@href}%
\providecommand \@@href[1]{\endgroup#1\@@endlink}%
\providecommand \@sanitize@url [0]{\catcode `\\12\catcode `\$12\catcode
  `\&12\catcode `\#12\catcode `\^12\catcode `\_12\catcode `\%12\relax}%
\providecommand \@@startlink[1]{}%
\providecommand \@@endlink[0]{}%
\providecommand \url  [0]{\begingroup\@sanitize@url \@url }%
\providecommand \@url [1]{\endgroup\@href {#1}{\urlprefix }}%
\providecommand \urlprefix  [0]{URL }%
\providecommand \Eprint [0]{\href }%
\providecommand \doibase [0]{http://dx.doi.org/}%
\providecommand \selectlanguage [0]{\@gobble}%
\providecommand \bibinfo  [0]{\@secondoftwo}%
\providecommand \bibfield  [0]{\@secondoftwo}%
\providecommand \translation [1]{[#1]}%
\providecommand \BibitemOpen [0]{}%
\providecommand \bibitemStop [0]{}%
\providecommand \bibitemNoStop [0]{.\EOS\space}%
\providecommand \EOS [0]{\spacefactor3000\relax}%
\providecommand \BibitemShut  [1]{\csname bibitem#1\endcsname}%
\let\auto@bib@innerbib\@empty
\bibitem [{\citenamefont {Rubin}\ \emph {et~al.}(1978)\citenamefont {Rubin},
  \citenamefont {Ford},\ and\ \citenamefont {Thonnard}}]{Rubin:1978kmz}%
  \BibitemOpen
  \bibfield  {author} {\bibinfo {author} {\bibfnamefont {V.~C.}\ \bibnamefont
  {Rubin}}, \bibinfo {author} {\bibfnamefont {W.~K.}\ \bibnamefont {Ford},
  \bibfnamefont {Jr.}}, \ and\ \bibinfo {author} {\bibfnamefont
  {N.}~\bibnamefont {Thonnard}},\ }\href {\doibase 10.1086/182804} {\bibfield
  {journal} {\bibinfo  {journal} {Astrophys. J.}\ }\textbf {\bibinfo {volume}
  {225}},\ \bibinfo {pages} {L107} (\bibinfo {year} {1978})}\BibitemShut
  {NoStop}%
\bibitem [{\citenamefont {Ade}\ \emph {et~al.}(2016)\citenamefont {Ade} \emph
  {et~al.}}]{Ade:2015xua}%
  \BibitemOpen
  \bibfield  {author} {\bibinfo {author} {\bibfnamefont {P.~A.~R.}\
  \bibnamefont {Ade}} \emph {et~al.} (\bibinfo {collaboration} {Planck}),\
  }\href {\doibase 10.1051/0004-6361/201525830} {\bibfield  {journal} {\bibinfo
   {journal} {Astron. Astrophys.}\ }\textbf {\bibinfo {volume} {594}},\
  \bibinfo {pages} {A13} (\bibinfo {year} {2016})},\ \Eprint
  {http://arxiv.org/abs/1502.01589} {arXiv:1502.01589 [astro-ph.CO]}
  \BibitemShut {NoStop}%
\bibitem [{\citenamefont
  {Feng}(2010)}]{doi:10.1146/annurev-astro-082708-101659}%
  \BibitemOpen
  \bibfield  {author} {\bibinfo {author} {\bibfnamefont {J.~L.}\ \bibnamefont
  {Feng}},\ }\href {\doibase 10.1146/annurev-astro-082708-101659} {\bibfield
  {journal} {\bibinfo  {journal} {Annual Review of Astronomy and Astrophysics}\
  }\textbf {\bibinfo {volume} {48}},\ \bibinfo {pages} {495} (\bibinfo {year}
  {2010})},\ \Eprint
  {http://arxiv.org/abs/https://doi.org/10.1146/annurev-astro-082708-101659}
  {https://doi.org/10.1146/annurev-astro-082708-101659} \BibitemShut {NoStop}%
\bibitem [{\citenamefont {{Buckley}}\ and\ \citenamefont
  {{Peter}}(2017)}]{2017arXiv171206615B}%
  \BibitemOpen
  \bibfield  {author} {\bibinfo {author} {\bibfnamefont {M.~R.}\ \bibnamefont
  {{Buckley}}}\ and\ \bibinfo {author} {\bibfnamefont {A.~H.~G.}\ \bibnamefont
  {{Peter}}},\ }\href@noop {} {\bibfield  {journal} {\bibinfo  {journal} {ArXiv
  e-prints}\ } (\bibinfo {year} {2017})},\ \Eprint
  {http://arxiv.org/abs/1712.06615} {arXiv:1712.06615} \BibitemShut {NoStop}%
\bibitem [{\citenamefont {{Patrignani}}\ \emph {et~al.}(2016)\citenamefont
  {{Patrignani}} \emph {et~al.}}]{2016ChPhC..40j0001P}%
  \BibitemOpen
  \bibfield  {author} {\bibinfo {author} {\bibfnamefont {C.}~\bibnamefont
  {{Patrignani}}} \emph {et~al.} (\bibinfo {collaboration} {Particle Data
  Group}),\ }\href {\doibase 10.1088/1674-1137/40/10/100001} {\bibfield
  {journal} {\bibinfo  {journal} {Chinese Physics C}\ }\textbf {\bibinfo
  {volume} {40}},\ \bibinfo {eid} {100001} (\bibinfo {year}
  {2016})}\BibitemShut {NoStop}%
\bibitem [{\citenamefont {{Freese}}(2017)}]{2017IJMPD..2630012F}%
  \BibitemOpen
  \bibfield  {author} {\bibinfo {author} {\bibfnamefont {K.}~\bibnamefont
  {{Freese}}},\ }\href {\doibase 10.1142/S0218271817300129} {\bibfield
  {journal} {\bibinfo  {journal} {International Journal of Modern Physics D}\
  }\textbf {\bibinfo {volume} {26}},\ \bibinfo {eid} {1730012-223} (\bibinfo
  {year} {2017})},\ \Eprint {http://arxiv.org/abs/1701.01840}
  {arXiv:1701.01840} \BibitemShut {NoStop}%
\bibitem [{\citenamefont {{Garrett}}\ and\ \citenamefont {{D{\=
  u}da}}(2011)}]{2011AdAst2011E...8G}%
  \BibitemOpen
  \bibfield  {author} {\bibinfo {author} {\bibfnamefont {K.}~\bibnamefont
  {{Garrett}}}\ and\ \bibinfo {author} {\bibfnamefont {G.}~\bibnamefont {{D{\=
  u}da}}},\ }\href {\doibase 10.1155/2011/968283} {\bibfield  {journal}
  {\bibinfo  {journal} {Advances in Astronomy}\ }\textbf {\bibinfo {volume}
  {2011}},\ \bibinfo {eid} {968283} (\bibinfo {year} {2011})},\ \Eprint
  {http://arxiv.org/abs/1006.2483} {arXiv:1006.2483 [hep-ph]} \BibitemShut
  {NoStop}%
\bibitem [{\citenamefont {Silk}\ \emph {et~al.}(1985)\citenamefont {Silk},
  \citenamefont {Olive},\ and\ \citenamefont {Srednicki}}]{PhysRevLett.55.257}%
  \BibitemOpen
  \bibfield  {author} {\bibinfo {author} {\bibfnamefont {J.}~\bibnamefont
  {Silk}}, \bibinfo {author} {\bibfnamefont {K.}~\bibnamefont {Olive}}, \ and\
  \bibinfo {author} {\bibfnamefont {M.}~\bibnamefont {Srednicki}},\ }\href
  {\doibase 10.1103/PhysRevLett.55.257} {\bibfield  {journal} {\bibinfo
  {journal} {Phys. Rev. Lett.}\ }\textbf {\bibinfo {volume} {55}},\ \bibinfo
  {pages} {257} (\bibinfo {year} {1985})}\BibitemShut {NoStop}%
\bibitem [{\citenamefont {{Peter}}(2009{\natexlab{a}})}]{2009PhRvD..79j3531P}%
  \BibitemOpen
  \bibfield  {author} {\bibinfo {author} {\bibfnamefont {A.~H.~G.}\
  \bibnamefont {{Peter}}},\ }\href {\doibase 10.1103/PhysRevD.79.103531}
  {\bibfield  {journal} {\bibinfo  {journal} {\prd}\ }\textbf {\bibinfo
  {volume} {79}},\ \bibinfo {eid} {103531} (\bibinfo {year}
  {2009}{\natexlab{a}})},\ \Eprint {http://arxiv.org/abs/0902.1344}
  {arXiv:0902.1344 [astro-ph.HE]} \BibitemShut {NoStop}%
\bibitem [{\citenamefont {{Widmark}}(2017)}]{2017JCAP...05..046W}%
  \BibitemOpen
  \bibfield  {author} {\bibinfo {author} {\bibfnamefont {A.}~\bibnamefont
  {{Widmark}}},\ }\href {\doibase 10.1088/1475-7516/2017/05/046} {\bibfield
  {journal} {\bibinfo  {journal} {Journal of Cosmology and Astroparticle
  Physics}\ }\textbf {\bibinfo {volume} {5}},\ \bibinfo {eid} {046} (\bibinfo
  {year} {2017})},\ \Eprint {http://arxiv.org/abs/1703.06878} {arXiv:1703.06878
  [hep-ph]} \BibitemShut {NoStop}%
\bibitem [{\citenamefont {Vincent}\ \emph {et~al.}(2015)\citenamefont
  {Vincent}, \citenamefont {Scott},\ and\ \citenamefont
  {Serenelli}}]{PhysRevLett.114.081302}%
  \BibitemOpen
  \bibfield  {author} {\bibinfo {author} {\bibfnamefont {A.~C.}\ \bibnamefont
  {Vincent}}, \bibinfo {author} {\bibfnamefont {P.}~\bibnamefont {Scott}}, \
  and\ \bibinfo {author} {\bibfnamefont {A.}~\bibnamefont {Serenelli}},\ }\href
  {\doibase 10.1103/PhysRevLett.114.081302} {\bibfield  {journal} {\bibinfo
  {journal} {Phys. Rev. Lett.}\ }\textbf {\bibinfo {volume} {114}},\ \bibinfo
  {pages} {081302} (\bibinfo {year} {2015})}\BibitemShut {NoStop}%
\bibitem [{\citenamefont {{Press}}\ and\ \citenamefont
  {{Spergel}}(1985)}]{1985ApJ...296..679P}%
  \BibitemOpen
  \bibfield  {author} {\bibinfo {author} {\bibfnamefont {W.~H.}\ \bibnamefont
  {{Press}}}\ and\ \bibinfo {author} {\bibfnamefont {D.~N.}\ \bibnamefont
  {{Spergel}}},\ }\href {\doibase 10.1086/163485} {\bibfield  {journal}
  {\bibinfo  {journal} {\apj}\ }\textbf {\bibinfo {volume} {296}},\ \bibinfo
  {pages} {679} (\bibinfo {year} {1985})}\BibitemShut {NoStop}%
\bibitem [{\citenamefont {Danninger}\ and\ \citenamefont
  {Rott}(2014)}]{Danninger:2014xza}%
  \BibitemOpen
  \bibfield  {author} {\bibinfo {author} {\bibfnamefont {M.}~\bibnamefont
  {Danninger}}\ and\ \bibinfo {author} {\bibfnamefont {C.}~\bibnamefont
  {Rott}},\ }\href {\doibase 10.1016/j.dark.2014.10.002} {\bibfield  {journal}
  {\bibinfo  {journal} {Phys. Dark Univ.}\ }\textbf {\bibinfo {volume} {5-6}},\
  \bibinfo {pages} {35} (\bibinfo {year} {2014})},\ \Eprint
  {http://arxiv.org/abs/1509.08230} {arXiv:1509.08230 [astro-ph.HE]}
  \BibitemShut {NoStop}%
\bibitem [{\citenamefont {Edsj{\"o}}(1995)}]{1995NuPhS..43..265E}%
  \BibitemOpen
  \bibfield  {author} {\bibinfo {author} {\bibfnamefont {J.}~\bibnamefont
  {Edsj{\"o}}},\ }\href {\doibase 10.1016/0920-5632(95)00487-T} {\bibfield
  {journal} {\bibinfo  {journal} {Nuclear Physics B Proceedings Supplements}\
  }\textbf {\bibinfo {volume} {43}},\ \bibinfo {pages} {265} (\bibinfo {year}
  {1995})},\ \Eprint {http://arxiv.org/abs/hep-ph/9504205} {hep-ph/9504205}
  \BibitemShut {NoStop}%
\bibitem [{\citenamefont {{Peter}}(2009{\natexlab{b}})}]{2009PhRvD..79j3532P}%
  \BibitemOpen
  \bibfield  {author} {\bibinfo {author} {\bibfnamefont {A.~H.~G.}\
  \bibnamefont {{Peter}}},\ }\href {\doibase 10.1103/PhysRevD.79.103532}
  {\bibfield  {journal} {\bibinfo  {journal} {\prd}\ }\textbf {\bibinfo
  {volume} {79}},\ \bibinfo {eid} {103532} (\bibinfo {year}
  {2009}{\natexlab{b}})},\ \Eprint {http://arxiv.org/abs/0902.1347}
  {arXiv:0902.1347 [astro-ph.HE]} \BibitemShut {NoStop}%
\bibitem [{\citenamefont {Bell}\ and\ \citenamefont
  {Petraki}(2011)}]{Bell:2011sn}%
  \BibitemOpen
  \bibfield  {author} {\bibinfo {author} {\bibfnamefont {N.~F.}\ \bibnamefont
  {Bell}}\ and\ \bibinfo {author} {\bibfnamefont {K.}~\bibnamefont {Petraki}},\
  }\href {\doibase 10.1088/1475-7516/2011/04/003} {\bibfield  {journal}
  {\bibinfo  {journal} {JCAP}\ }\textbf {\bibinfo {volume} {1104}},\ \bibinfo
  {pages} {003} (\bibinfo {year} {2011})},\ \Eprint
  {http://arxiv.org/abs/1102.2958} {arXiv:1102.2958 [hep-ph]} \BibitemShut
  {NoStop}%
\bibitem [{\citenamefont {Feng}\ \emph
  {et~al.}(2016{\natexlab{a}})\citenamefont {Feng}, \citenamefont {Smolinsky},\
  and\ \citenamefont {Tanedo}}]{Feng:2016ijc}%
  \BibitemOpen
  \bibfield  {author} {\bibinfo {author} {\bibfnamefont {J.~L.}\ \bibnamefont
  {Feng}}, \bibinfo {author} {\bibfnamefont {J.}~\bibnamefont {Smolinsky}}, \
  and\ \bibinfo {author} {\bibfnamefont {P.}~\bibnamefont {Tanedo}},\ }\href
  {\doibase 10.1103/PhysRevD.93.115036, 10.1103/PhysRevD.96.099903} {\bibfield
  {journal} {\bibinfo  {journal} {Phys. Rev.}\ }\textbf {\bibinfo {volume}
  {D93}},\ \bibinfo {pages} {115036} (\bibinfo {year} {2016}{\natexlab{a}})},\
  \bibinfo {note} {[Erratum: Phys. Rev.D96,(2017)]},\ \Eprint
  {http://arxiv.org/abs/1602.01465} {arXiv:1602.01465 [hep-ph]} \BibitemShut
  {NoStop}%
\bibitem [{\citenamefont {Kouvaris}\ \emph {et~al.}(2016)\citenamefont
  {Kouvaris}, \citenamefont {Langæble},\ and\ \citenamefont
  {Nielsen}}]{Kouvaris:2016ltf}%
  \BibitemOpen
  \bibfield  {author} {\bibinfo {author} {\bibfnamefont {C.}~\bibnamefont
  {Kouvaris}}, \bibinfo {author} {\bibfnamefont {K.}~\bibnamefont {Langæble}},
  \ and\ \bibinfo {author} {\bibfnamefont {N.~G.}\ \bibnamefont {Nielsen}},\
  }\href {\doibase 10.1088/1475-7516/2016/10/012} {\bibfield  {journal}
  {\bibinfo  {journal} {JCAP}\ }\textbf {\bibinfo {volume} {1610}},\ \bibinfo
  {pages} {012} (\bibinfo {year} {2016})},\ \Eprint
  {http://arxiv.org/abs/1607.00374} {arXiv:1607.00374 [hep-ph]} \BibitemShut
  {NoStop}%
\bibitem [{\citenamefont {Akerib}\ \emph
  {et~al.}(2017{\natexlab{a}})\citenamefont {Akerib} \emph
  {et~al.}}]{PhysRevLett.118.021303}%
  \BibitemOpen
  \bibfield  {author} {\bibinfo {author} {\bibfnamefont {D.~S.}\ \bibnamefont
  {Akerib}} \emph {et~al.} (\bibinfo {collaboration} {LUX Collaboration}),\
  }\href {\doibase 10.1103/PhysRevLett.118.021303} {\bibfield  {journal}
  {\bibinfo  {journal} {Phys. Rev. Lett.}\ }\textbf {\bibinfo {volume} {118}},\
  \bibinfo {pages} {021303} (\bibinfo {year} {2017}{\natexlab{a}})}\BibitemShut
  {NoStop}%
\bibitem [{\citenamefont {{Liu}}\ \emph {et~al.}(2017)\citenamefont {{Liu}},
  \citenamefont {{Chen}},\ and\ \citenamefont {{Ji}}}]{2017NatPh..13..212L}%
  \BibitemOpen
  \bibfield  {author} {\bibinfo {author} {\bibfnamefont {J.}~\bibnamefont
  {{Liu}}}, \bibinfo {author} {\bibfnamefont {X.}~\bibnamefont {{Chen}}}, \
  and\ \bibinfo {author} {\bibfnamefont {X.}~\bibnamefont {{Ji}}},\ }\href
  {\doibase 10.1038/nphys4039} {\bibfield  {journal} {\bibinfo  {journal}
  {Nature Physics}\ }\textbf {\bibinfo {volume} {13}},\ \bibinfo {pages} {212}
  (\bibinfo {year} {2017})},\ \Eprint {http://arxiv.org/abs/1709.00688}
  {arXiv:1709.00688} \BibitemShut {NoStop}%
\bibitem [{\citenamefont {{Marrod{\'a}n Undagoitia}}\ and\ \citenamefont
  {{Rauch}}(2016)}]{2016JPhG...43a3001M}%
  \BibitemOpen
  \bibfield  {author} {\bibinfo {author} {\bibfnamefont {T.}~\bibnamefont
  {{Marrod{\'a}n Undagoitia}}}\ and\ \bibinfo {author} {\bibfnamefont
  {L.}~\bibnamefont {{Rauch}}},\ }\href {\doibase
  10.1088/0954-3899/43/1/013001} {\bibfield  {journal} {\bibinfo  {journal}
  {Journal of Physics G Nuclear Physics}\ }\textbf {\bibinfo {volume} {43}},\
  \bibinfo {eid} {013001} (\bibinfo {year} {2016})},\ \Eprint
  {http://arxiv.org/abs/1509.08767} {arXiv:1509.08767 [physics.ins-det]}
  \BibitemShut {NoStop}%
\bibitem [{\citenamefont {Cui}\ \emph {et~al.}(2017)\citenamefont {Cui} \emph
  {et~al.}}]{PhysRevLett.119.181302}%
  \BibitemOpen
  \bibfield  {author} {\bibinfo {author} {\bibfnamefont {X.}~\bibnamefont
  {Cui}} \emph {et~al.} (\bibinfo {collaboration} {PandaX-II Collaboration}),\
  }\href {\doibase 10.1103/PhysRevLett.119.181302} {\bibfield  {journal}
  {\bibinfo  {journal} {Phys. Rev. Lett.}\ }\textbf {\bibinfo {volume} {119}},\
  \bibinfo {pages} {181302} (\bibinfo {year} {2017})}\BibitemShut {NoStop}%
\bibitem [{\citenamefont {{Fu}}\ \emph {et~al.}(2017)\citenamefont {{Fu}} \emph
  {et~al.}}]{2017PhRvL.118g1301F}%
  \BibitemOpen
  \bibfield  {author} {\bibinfo {author} {\bibfnamefont {C.}~\bibnamefont
  {{Fu}}} \emph {et~al.},\ }\href {\doibase 10.1103/PhysRevLett.118.071301}
  {\bibfield  {journal} {\bibinfo  {journal} {Physical Review Letters}\
  }\textbf {\bibinfo {volume} {118}},\ \bibinfo {eid} {071301} (\bibinfo {year}
  {2017})},\ \Eprint {http://arxiv.org/abs/1611.06553} {arXiv:1611.06553
  [hep-ex]} \BibitemShut {NoStop}%
\bibitem [{\citenamefont {{Amole}}\ \emph {et~al.}(2017)\citenamefont {{Amole}}
  \emph {et~al.}}]{2017PhRvL.118y1301A}%
  \BibitemOpen
  \bibfield  {author} {\bibinfo {author} {\bibfnamefont {C.}~\bibnamefont
  {{Amole}}} \emph {et~al.} (\bibinfo {collaboration} {PICO Collaboration}),\
  }\href {\doibase 10.1103/PhysRevLett.118.251301} {\bibfield  {journal}
  {\bibinfo  {journal} {Physical Review Letters}\ }\textbf {\bibinfo {volume}
  {118}},\ \bibinfo {eid} {251301} (\bibinfo {year} {2017})},\ \Eprint
  {http://arxiv.org/abs/1702.07666} {arXiv:1702.07666} \BibitemShut {NoStop}%
\bibitem [{\citenamefont {Xia}\ \emph {et~al.}(2018)\citenamefont {Xia} \emph
  {et~al.}}]{1681014}%
  \BibitemOpen
  \bibfield  {author} {\bibinfo {author} {\bibfnamefont {J.}~\bibnamefont
  {Xia}} \emph {et~al.},\ }\href@noop {} {\bibfield  {journal} {\bibinfo
  {journal} {ArXiv e-prints}\ } (\bibinfo {year} {2018})},\ \Eprint
  {http://arxiv.org/abs/1807.01936} {arXiv:1807.01936 [hep-ex]} \BibitemShut
  {NoStop}%
\bibitem [{\citenamefont {Akerib}\ \emph
  {et~al.}(2017{\natexlab{b}})\citenamefont {Akerib} \emph
  {et~al.}}]{Akerib:2017kat}%
  \BibitemOpen
  \bibfield  {author} {\bibinfo {author} {\bibfnamefont {D.~S.}\ \bibnamefont
  {Akerib}} \emph {et~al.} (\bibinfo {collaboration} {LUX}),\ }\href {\doibase
  10.1103/PhysRevLett.118.251302} {\bibfield  {journal} {\bibinfo  {journal}
  {Phys. Rev. Lett.}\ }\textbf {\bibinfo {volume} {118}},\ \bibinfo {pages}
  {251302} (\bibinfo {year} {2017}{\natexlab{b}})},\ \Eprint
  {http://arxiv.org/abs/1705.03380} {arXiv:1705.03380 [astro-ph.CO]}
  \BibitemShut {NoStop}%
\bibitem [{\citenamefont {Aprile}\ \emph {et~al.}(2017)\citenamefont {Aprile}
  \emph {et~al.}}]{PhysRevLett.119.181301}%
  \BibitemOpen
  \bibfield  {author} {\bibinfo {author} {\bibfnamefont {E.}~\bibnamefont
  {Aprile}} \emph {et~al.} (\bibinfo {collaboration} {XENON Collaboration}),\
  }\href {\doibase 10.1103/PhysRevLett.119.181301} {\bibfield  {journal}
  {\bibinfo  {journal} {Phys. Rev. Lett.}\ }\textbf {\bibinfo {volume} {119}},\
  \bibinfo {pages} {181301} (\bibinfo {year} {2017})}\BibitemShut {NoStop}%
\bibitem [{\citenamefont {Aprile}\ \emph {et~al.}(2018)\citenamefont {Aprile}
  \emph {et~al.}}]{Aprile:2018dbl}%
  \BibitemOpen
  \bibfield  {author} {\bibinfo {author} {\bibfnamefont {E.}~\bibnamefont
  {Aprile}} \emph {et~al.} (\bibinfo {collaboration} {XENON}),\ }\href
  {\doibase 10.1103/PhysRevLett.121.111302} {\bibfield  {journal} {\bibinfo
  {journal} {Physical Review Letters}\ ,\ \bibinfo {pages} {111302}} (\bibinfo
  {year} {2018})},\ \Eprint {http://arxiv.org/abs/1805.12562} {arXiv:1805.12562
  [astro-ph.CO]} \BibitemShut {NoStop}%
\bibitem [{\citenamefont {Conrad}\ and\ \citenamefont
  {Reimer}(2017)}]{Indirectdm}%
  \BibitemOpen
  \bibfield  {author} {\bibinfo {author} {\bibfnamefont {J.}~\bibnamefont
  {Conrad}}\ and\ \bibinfo {author} {\bibfnamefont {O.}~\bibnamefont
  {Reimer}},\ }\href {http://dx.doi.org/10.1038/nphys4049} {\bibfield
  {journal} {\bibinfo  {journal} {Nat Phys}\ }\textbf {\bibinfo {volume}
  {13}},\ \bibinfo {pages} {224} (\bibinfo {year} {2017})}\BibitemShut
  {NoStop}%
\bibitem [{\citenamefont {Leane}\ \emph {et~al.}(2018)\citenamefont {Leane},
  \citenamefont {Slatyer}, \citenamefont {Beacom},\ and\ \citenamefont
  {Ng}}]{Leane:2018kjk}%
  \BibitemOpen
  \bibfield  {author} {\bibinfo {author} {\bibfnamefont {R.~K.}\ \bibnamefont
  {Leane}}, \bibinfo {author} {\bibfnamefont {T.~R.}\ \bibnamefont {Slatyer}},
  \bibinfo {author} {\bibfnamefont {J.~F.}\ \bibnamefont {Beacom}}, \ and\
  \bibinfo {author} {\bibfnamefont {K.~C.~Y.}\ \bibnamefont {Ng}},\ }\href
  {\doibase 10.1103/PhysRevD.98.023016} {\bibfield  {journal} {\bibinfo
  {journal} {\prd}\ }\textbf {\bibinfo {volume} {98}},\ \bibinfo {eid} {023016}
  (\bibinfo {year} {2018})},\ \Eprint {http://arxiv.org/abs/1805.10305}
  {arXiv:1805.10305 [hep-ph]} \BibitemShut {NoStop}%
\bibitem [{\citenamefont {{Aartsen}}\ \emph {et~al.}(2017)\citenamefont
  {{Aartsen}} \emph {et~al.}}]{2017EPJC...77..146A}%
  \BibitemOpen
  \bibfield  {author} {\bibinfo {author} {\bibfnamefont {M.~G.}\ \bibnamefont
  {{Aartsen}}} \emph {et~al.},\ }\href {\doibase
  10.1140/epjc/s10052-017-4689-9} {\bibfield  {journal} {\bibinfo  {journal}
  {European Physical Journal C}\ }\textbf {\bibinfo {volume} {77}},\ \bibinfo
  {eid} {146} (\bibinfo {year} {2017})},\ \Eprint
  {http://arxiv.org/abs/1612.05949} {arXiv:1612.05949 [astro-ph.HE]}
  \BibitemShut {NoStop}%
\bibitem [{\citenamefont {{Adrian-Martinez}}\ \emph {et~al.}(2016)\citenamefont
  {{Adrian-Martinez}} \emph {et~al.}}]{2016PhLB..759...69A}%
  \BibitemOpen
  \bibfield  {author} {\bibinfo {author} {\bibfnamefont {S.}~\bibnamefont
  {{Adrian-Martinez}}} \emph {et~al.},\ }\href {\doibase
  10.1016/j.physletb.2016.05.019} {\bibfield  {journal} {\bibinfo  {journal}
  {Physics Letters B}\ }\textbf {\bibinfo {volume} {759}},\ \bibinfo {pages}
  {69} (\bibinfo {year} {2016})},\ \Eprint {http://arxiv.org/abs/1603.02228}
  {arXiv:1603.02228 [astro-ph.HE]} \BibitemShut {NoStop}%
\bibitem [{\citenamefont {Choi}\ \emph {et~al.}(2015)\citenamefont {Choi} \emph
  {et~al.}}]{Choi:2015ara}%
  \BibitemOpen
  \bibfield  {author} {\bibinfo {author} {\bibfnamefont {K.}~\bibnamefont
  {Choi}} \emph {et~al.} (\bibinfo {collaboration} {Super-Kamiokande}),\ }\href
  {\doibase 10.1103/PhysRevLett.114.141301} {\bibfield  {journal} {\bibinfo
  {journal} {Phys. Rev. Lett.}\ }\textbf {\bibinfo {volume} {114}},\ \bibinfo
  {pages} {141301} (\bibinfo {year} {2015})},\ \Eprint
  {http://arxiv.org/abs/1503.04858} {arXiv:1503.04858 [hep-ex]} \BibitemShut
  {NoStop}%
\bibitem [{\citenamefont {{Leane}}\ \emph {et~al.}(2017)\citenamefont
  {{Leane}}, \citenamefont {{Ng}},\ and\ \citenamefont
  {{Beacom}}}]{2017PhRvD..95l3016L}%
  \BibitemOpen
  \bibfield  {author} {\bibinfo {author} {\bibfnamefont {R.~K.}\ \bibnamefont
  {{Leane}}}, \bibinfo {author} {\bibfnamefont {K.~C.~Y.}\ \bibnamefont
  {{Ng}}}, \ and\ \bibinfo {author} {\bibfnamefont {J.~F.}\ \bibnamefont
  {{Beacom}}},\ }\href {\doibase 10.1103/PhysRevD.95.123016} {\bibfield
  {journal} {\bibinfo  {journal} {Phys. Rev. D}\ }\textbf {\bibinfo {volume}
  {95}},\ \bibinfo {eid} {123016} (\bibinfo {year} {2017})},\ \Eprint
  {http://arxiv.org/abs/1703.04629} {arXiv:1703.04629 [astro-ph.HE]}
  \BibitemShut {NoStop}%
\bibitem [{\citenamefont {{Hooper}}(2001)}]{2001hep.ph....3277H}%
  \BibitemOpen
  \bibfield  {author} {\bibinfo {author} {\bibfnamefont {D.~W.}\ \bibnamefont
  {{Hooper}}},\ }\href@noop {} {\bibfield  {journal} {\bibinfo  {journal}
  {ArxIv e-prints}\ } (\bibinfo {year} {2001})},\ \Eprint
  {http://arxiv.org/abs/hep-ph/0103277} {hep-ph/0103277} \BibitemShut {NoStop}%
\bibitem [{\citenamefont {{Sivertsson}}\ and\ \citenamefont
  {{Edsj{\"o}}}(2010)}]{2010PhRvD..81f3502S}%
  \BibitemOpen
  \bibfield  {author} {\bibinfo {author} {\bibfnamefont {S.}~\bibnamefont
  {{Sivertsson}}}\ and\ \bibinfo {author} {\bibfnamefont {J.}~\bibnamefont
  {{Edsj{\"o}}}},\ }\href {\doibase 10.1103/PhysRevD.81.063502} {\bibfield
  {journal} {\bibinfo  {journal} {\prd}\ }\textbf {\bibinfo {volume} {81}},\
  \bibinfo {eid} {063502} (\bibinfo {year} {2010})},\ \Eprint
  {http://arxiv.org/abs/0910.0017} {arXiv:0910.0017 [astro-ph.HE]} \BibitemShut
  {NoStop}%
\bibitem [{\citenamefont {{Edsj{\"o}}}(1997)}]{1997PhDT.........5E}%
  \BibitemOpen
  \bibfield  {author} {\bibinfo {author} {\bibfnamefont {J.}~\bibnamefont
  {{Edsj{\"o}}}},\ }\emph {\bibinfo {title} {{Aspects of Neutrino Detection of
  Neutralino Dark Matter}}},\ \href@noop {} {Ph.D. thesis},\ \bibinfo  {school}
  {Uppsala Univ.~(preprint hep-ph/9704384)} (\bibinfo {year}
  {1997})\BibitemShut {NoStop}%
\bibitem [{\citenamefont {{Schuster}}\ \emph
  {et~al.}(2010{\natexlab{a}})\citenamefont {{Schuster}}, \citenamefont
  {{Toro}}, \citenamefont {{Weiner}},\ and\ \citenamefont
  {{Yavin}}}]{2010PhRvD..82k5012S}%
  \BibitemOpen
  \bibfield  {author} {\bibinfo {author} {\bibfnamefont {P.}~\bibnamefont
  {{Schuster}}}, \bibinfo {author} {\bibfnamefont {N.}~\bibnamefont {{Toro}}},
  \bibinfo {author} {\bibfnamefont {N.}~\bibnamefont {{Weiner}}}, \ and\
  \bibinfo {author} {\bibfnamefont {I.}~\bibnamefont {{Yavin}}},\ }\href
  {\doibase 10.1103/PhysRevD.82.115012} {\bibfield  {journal} {\bibinfo
  {journal} {\prd}\ }\textbf {\bibinfo {volume} {82}},\ \bibinfo {eid} {115012}
  (\bibinfo {year} {2010}{\natexlab{a}})},\ \Eprint
  {http://arxiv.org/abs/0910.1839} {arXiv:0910.1839 [hep-ph]} \BibitemShut
  {NoStop}%
\bibitem [{\citenamefont {{Schuster}}\ \emph
  {et~al.}(2010{\natexlab{b}})\citenamefont {{Schuster}}, \citenamefont
  {{Toro}},\ and\ \citenamefont {{Yavin}}}]{2010PhRvD..81a6002S}%
  \BibitemOpen
  \bibfield  {author} {\bibinfo {author} {\bibfnamefont {P.}~\bibnamefont
  {{Schuster}}}, \bibinfo {author} {\bibfnamefont {N.}~\bibnamefont {{Toro}}},
  \ and\ \bibinfo {author} {\bibfnamefont {I.}~\bibnamefont {{Yavin}}},\ }\href
  {\doibase 10.1103/PhysRevD.81.016002} {\bibfield  {journal} {\bibinfo
  {journal} {\prd}\ }\textbf {\bibinfo {volume} {81}},\ \bibinfo {eid} {016002}
  (\bibinfo {year} {2010}{\natexlab{b}})},\ \Eprint
  {http://arxiv.org/abs/0910.1602} {arXiv:0910.1602 [hep-ph]} \BibitemShut
  {NoStop}%
\bibitem [{\citenamefont {{Batell}}\ \emph {et~al.}(2010)\citenamefont
  {{Batell}}, \citenamefont {{Pospelov}}, \citenamefont {{Ritz}},\ and\
  \citenamefont {{Shang}}}]{2010PhRvD..81g5004B}%
  \BibitemOpen
  \bibfield  {author} {\bibinfo {author} {\bibfnamefont {B.}~\bibnamefont
  {{Batell}}}, \bibinfo {author} {\bibfnamefont {M.}~\bibnamefont
  {{Pospelov}}}, \bibinfo {author} {\bibfnamefont {A.}~\bibnamefont {{Ritz}}},
  \ and\ \bibinfo {author} {\bibfnamefont {Y.}~\bibnamefont {{Shang}}},\ }\href
  {\doibase 10.1103/PhysRevD.81.075004} {\bibfield  {journal} {\bibinfo
  {journal} {Phys. Rev. D}\ }\textbf {\bibinfo {volume} {81}},\ \bibinfo {eid}
  {075004} (\bibinfo {year} {2010})},\ \Eprint {http://arxiv.org/abs/0910.1567}
  {arXiv:0910.1567 [hep-ph]} \BibitemShut {NoStop}%
\bibitem [{\citenamefont {Curtin}\ \emph {et~al.}(2018)\citenamefont {Curtin}
  \emph {et~al.}}]{Curtin:2018mvb}%
  \BibitemOpen
  \bibfield  {author} {\bibinfo {author} {\bibfnamefont {D.}~\bibnamefont
  {Curtin}} \emph {et~al.},\ }\href@noop {} {\bibfield  {journal} {\bibinfo
  {journal} {ArXiv e-prints}\ } (\bibinfo {year} {2018})},\ \Eprint
  {http://arxiv.org/abs/1806.07396} {arXiv:1806.07396 [hep-ph]} \BibitemShut
  {NoStop}%
\bibitem [{\citenamefont {Holdom}(1986)}]{Holdom:1985ag}%
  \BibitemOpen
  \bibfield  {author} {\bibinfo {author} {\bibfnamefont {B.}~\bibnamefont
  {Holdom}},\ }\href {\doibase 10.1016/0370-2693(86)91377-8} {\bibfield
  {journal} {\bibinfo  {journal} {Phys. Lett.}\ }\textbf {\bibinfo {volume}
  {166B}},\ \bibinfo {pages} {196} (\bibinfo {year} {1986})}\BibitemShut
  {NoStop}%
\bibitem [{\citenamefont {{Martin}}(1998)}]{1998pesu.book....1M}%
  \BibitemOpen
  \bibfield  {author} {\bibinfo {author} {\bibfnamefont {S.~P.}\ \bibnamefont
  {{Martin}}},\ }\enquote {\bibinfo {title} {{A Supersymmetry Primer}},}\ in\
  \href {\doibase 10.1142/9789812839657_0001} {\emph {\bibinfo {booktitle}
  {Perspectives on Supersymmetry}}},\ \bibinfo {editor} {edited by\ \bibinfo
  {editor} {\bibfnamefont {G.~L.}\ \bibnamefont {{Kane}}}}\ (\bibinfo
  {publisher} {World Scientific Publishing Co},\ \bibinfo {year} {1998})\ pp.\
  \bibinfo {pages} {1--98}\BibitemShut {NoStop}%
\bibitem [{\citenamefont {Pospelov}\ \emph {et~al.}(2008)\citenamefont
  {Pospelov}, \citenamefont {Ritz},\ and\ \citenamefont
  {Voloshin}}]{Pospelov:2007mp}%
  \BibitemOpen
  \bibfield  {author} {\bibinfo {author} {\bibfnamefont {M.}~\bibnamefont
  {Pospelov}}, \bibinfo {author} {\bibfnamefont {A.}~\bibnamefont {Ritz}}, \
  and\ \bibinfo {author} {\bibfnamefont {M.~B.}\ \bibnamefont {Voloshin}},\
  }\href {\doibase 10.1016/j.physletb.2008.02.052} {\bibfield  {journal}
  {\bibinfo  {journal} {Phys. Lett.}\ }\textbf {\bibinfo {volume} {B662}},\
  \bibinfo {pages} {53} (\bibinfo {year} {2008})},\ \Eprint
  {http://arxiv.org/abs/0711.4866} {arXiv:0711.4866 [hep-ph]} \BibitemShut
  {NoStop}%
\bibitem [{\citenamefont {Pospelov}\ and\ \citenamefont
  {Ritz}(2009)}]{Pospelov:2008jd}%
  \BibitemOpen
  \bibfield  {author} {\bibinfo {author} {\bibfnamefont {M.}~\bibnamefont
  {Pospelov}}\ and\ \bibinfo {author} {\bibfnamefont {A.}~\bibnamefont
  {Ritz}},\ }\href {\doibase 10.1016/j.physletb.2008.12.012} {\bibfield
  {journal} {\bibinfo  {journal} {Phys. Lett.}\ }\textbf {\bibinfo {volume}
  {B671}},\ \bibinfo {pages} {391} (\bibinfo {year} {2009})},\ \Eprint
  {http://arxiv.org/abs/0810.1502} {arXiv:0810.1502 [hep-ph]} \BibitemShut
  {NoStop}%
\bibitem [{\citenamefont {Batell}\ \emph {et~al.}(2010)\citenamefont {Batell},
  \citenamefont {Pospelov}, \citenamefont {Ritz},\ and\ \citenamefont
  {Shang}}]{Batell:2009zp}%
  \BibitemOpen
  \bibfield  {author} {\bibinfo {author} {\bibfnamefont {B.}~\bibnamefont
  {Batell}}, \bibinfo {author} {\bibfnamefont {M.}~\bibnamefont {Pospelov}},
  \bibinfo {author} {\bibfnamefont {A.}~\bibnamefont {Ritz}}, \ and\ \bibinfo
  {author} {\bibfnamefont {Y.}~\bibnamefont {Shang}},\ }\href {\doibase
  10.1103/PhysRevD.81.075004} {\bibfield  {journal} {\bibinfo  {journal} {Phys.
  Rev.}\ }\textbf {\bibinfo {volume} {D81}},\ \bibinfo {pages} {075004}
  (\bibinfo {year} {2010})},\ \Eprint {http://arxiv.org/abs/0910.1567}
  {arXiv:0910.1567 [hep-ph]} \BibitemShut {NoStop}%
\bibitem [{\citenamefont {Rothstein}\ \emph {et~al.}(2009)\citenamefont
  {Rothstein}, \citenamefont {Schwetz},\ and\ \citenamefont
  {Zupan}}]{Rothstein:2009pm}%
  \BibitemOpen
  \bibfield  {author} {\bibinfo {author} {\bibfnamefont {I.~Z.}\ \bibnamefont
  {Rothstein}}, \bibinfo {author} {\bibfnamefont {T.}~\bibnamefont {Schwetz}},
  \ and\ \bibinfo {author} {\bibfnamefont {J.}~\bibnamefont {Zupan}},\ }\href
  {\doibase 10.1088/1475-7516/2009/07/018} {\bibfield  {journal} {\bibinfo
  {journal} {JCAP}\ }\textbf {\bibinfo {volume} {0907}},\ \bibinfo {pages}
  {018} (\bibinfo {year} {2009})},\ \Eprint {http://arxiv.org/abs/0903.3116}
  {arXiv:0903.3116 [astro-ph.HE]} \BibitemShut {NoStop}%
\bibitem [{\citenamefont {Chen}\ \emph {et~al.}(2009)\citenamefont {Chen},
  \citenamefont {Cline},\ and\ \citenamefont {Frey}}]{Chen:2009ab}%
  \BibitemOpen
  \bibfield  {author} {\bibinfo {author} {\bibfnamefont {F.}~\bibnamefont
  {Chen}}, \bibinfo {author} {\bibfnamefont {J.~M.}\ \bibnamefont {Cline}}, \
  and\ \bibinfo {author} {\bibfnamefont {A.~R.}\ \bibnamefont {Frey}},\ }\href
  {\doibase 10.1103/PhysRevD.80.083516} {\bibfield  {journal} {\bibinfo
  {journal} {Phys. Rev.}\ }\textbf {\bibinfo {volume} {D80}},\ \bibinfo {pages}
  {083516} (\bibinfo {year} {2009})},\ \Eprint {http://arxiv.org/abs/0907.4746}
  {arXiv:0907.4746 [hep-ph]} \BibitemShut {NoStop}%
\bibitem [{\citenamefont {Meade}\ \emph {et~al.}(2010)\citenamefont {Meade},
  \citenamefont {Nussinov}, \citenamefont {Papucci},\ and\ \citenamefont
  {Volansky}}]{Meade:2009mu}%
  \BibitemOpen
  \bibfield  {author} {\bibinfo {author} {\bibfnamefont {P.}~\bibnamefont
  {Meade}}, \bibinfo {author} {\bibfnamefont {S.}~\bibnamefont {Nussinov}},
  \bibinfo {author} {\bibfnamefont {M.}~\bibnamefont {Papucci}}, \ and\
  \bibinfo {author} {\bibfnamefont {T.}~\bibnamefont {Volansky}},\ }\href
  {\doibase 10.1007/JHEP06(2010)029} {\bibfield  {journal} {\bibinfo  {journal}
  {JHEP}\ }\textbf {\bibinfo {volume} {06}},\ \bibinfo {pages} {029} (\bibinfo
  {year} {2010})},\ \Eprint {http://arxiv.org/abs/0910.4160} {arXiv:0910.4160
  [hep-ph]} \BibitemShut {NoStop}%
\bibitem [{\citenamefont {Feng}\ \emph
  {et~al.}(2016{\natexlab{b}})\citenamefont {Feng}, \citenamefont {Smolinsky},\
  and\ \citenamefont {Tanedo}}]{Feng:2015hja}%
  \BibitemOpen
  \bibfield  {author} {\bibinfo {author} {\bibfnamefont {J.~L.}\ \bibnamefont
  {Feng}}, \bibinfo {author} {\bibfnamefont {J.}~\bibnamefont {Smolinsky}}, \
  and\ \bibinfo {author} {\bibfnamefont {P.}~\bibnamefont {Tanedo}},\ }\href
  {\doibase 10.1103/PhysRevD.93.015014} {\bibfield  {journal} {\bibinfo
  {journal} {Phys. Rev.}\ }\textbf {\bibinfo {volume} {D93}},\ \bibinfo {pages}
  {015014} (\bibinfo {year} {2016}{\natexlab{b}})},\ \Eprint
  {http://arxiv.org/abs/1509.07525} {arXiv:1509.07525 [hep-ph]} \BibitemShut
  {NoStop}%
\bibitem [{\citenamefont {Bell}\ \emph {et~al.}(2016)\citenamefont {Bell},
  \citenamefont {Cai},\ and\ \citenamefont {Leane}}]{Bell:2016fqf}%
  \BibitemOpen
  \bibfield  {author} {\bibinfo {author} {\bibfnamefont {N.~F.}\ \bibnamefont
  {Bell}}, \bibinfo {author} {\bibfnamefont {Y.}~\bibnamefont {Cai}}, \ and\
  \bibinfo {author} {\bibfnamefont {R.~K.}\ \bibnamefont {Leane}},\ }\href
  {\doibase 10.1088/1475-7516/2016/08/001} {\bibfield  {journal} {\bibinfo
  {journal} {JCAP}\ }\textbf {\bibinfo {volume} {1608}},\ \bibinfo {pages}
  {001} (\bibinfo {year} {2016})},\ \Eprint {http://arxiv.org/abs/1605.09382}
  {arXiv:1605.09382 [hep-ph]} \BibitemShut {NoStop}%
\bibitem [{\citenamefont {{Bell}}\ \emph {et~al.}(2017)\citenamefont {{Bell}},
  \citenamefont {{Cai}},\ and\ \citenamefont {{Leane}}}]{Bell:2016uhg}%
  \BibitemOpen
  \bibfield  {author} {\bibinfo {author} {\bibfnamefont {N.~F.}\ \bibnamefont
  {{Bell}}}, \bibinfo {author} {\bibfnamefont {Y.}~\bibnamefont {{Cai}}}, \
  and\ \bibinfo {author} {\bibfnamefont {R.~K.}\ \bibnamefont {{Leane}}},\
  }\href {\doibase 10.1088/1475-7516/2017/01/039} {\bibfield  {journal}
  {\bibinfo  {journal} {Journal of Cosmology and Astroparticle Physics}\
  }\textbf {\bibinfo {volume} {1}},\ \bibinfo {eid} {039} (\bibinfo {year}
  {2017})},\ \Eprint {http://arxiv.org/abs/1610.03063} {arXiv:1610.03063
  [hep-ph]} \BibitemShut {NoStop}%
\bibitem [{\citenamefont {Adrian-Martinez}\ \emph {et~al.}(2016)\citenamefont
  {Adrian-Martinez} \emph {et~al.}}]{Adrian-Martinez:2016ujo}%
  \BibitemOpen
  \bibfield  {author} {\bibinfo {author} {\bibfnamefont {S.}~\bibnamefont
  {Adrian-Martinez}} \emph {et~al.} (\bibinfo {collaboration} {ANTARES}),\
  }\href {\doibase 10.1088/1475-7516/2016/05/016} {\bibfield  {journal}
  {\bibinfo  {journal} {JCAP}\ }\textbf {\bibinfo {volume} {1605}},\ \bibinfo
  {pages} {016} (\bibinfo {year} {2016})},\ \Eprint
  {http://arxiv.org/abs/1602.07000} {arXiv:1602.07000 [hep-ex]} \BibitemShut
  {NoStop}%
\bibitem [{\citenamefont {Allahverdi}\ \emph {et~al.}(2017)\citenamefont
  {Allahverdi}, \citenamefont {Gao}, \citenamefont {Knockel},\ and\
  \citenamefont {Shalgar}}]{Allahverdi:2016fvl}%
  \BibitemOpen
  \bibfield  {author} {\bibinfo {author} {\bibfnamefont {R.}~\bibnamefont
  {Allahverdi}}, \bibinfo {author} {\bibfnamefont {Y.}~\bibnamefont {Gao}},
  \bibinfo {author} {\bibfnamefont {B.}~\bibnamefont {Knockel}}, \ and\
  \bibinfo {author} {\bibfnamefont {S.}~\bibnamefont {Shalgar}},\ }\href
  {\doibase 10.1103/PhysRevD.95.075001} {\bibfield  {journal} {\bibinfo
  {journal} {Phys. Rev. D}\ }\textbf {\bibinfo {volume} {95}},\ \bibinfo
  {pages} {075001} (\bibinfo {year} {2017})}\BibitemShut {NoStop}%
\bibitem [{\citenamefont {{Ardid}}\ \emph {et~al.}(2017)\citenamefont
  {{Ardid}}, \citenamefont {{Felis}}, \citenamefont {{Herrero}},\ and\
  \citenamefont {{Mart{\'{\i}}nez-Mora}}}]{Ardid:2017lry}%
  \BibitemOpen
  \bibfield  {author} {\bibinfo {author} {\bibfnamefont {M.}~\bibnamefont
  {{Ardid}}}, \bibinfo {author} {\bibfnamefont {I.}~\bibnamefont {{Felis}}},
  \bibinfo {author} {\bibfnamefont {A.}~\bibnamefont {{Herrero}}}, \ and\
  \bibinfo {author} {\bibfnamefont {J.~A.}\ \bibnamefont
  {{Mart{\'{\i}}nez-Mora}}},\ }\href {\doibase 10.1088/1475-7516/2017/04/010}
  {\bibfield  {journal} {\bibinfo  {journal} {Journal of Cosmology and
  Astroparticle Physics}\ }\textbf {\bibinfo {volume} {4}},\ \bibinfo {eid}
  {010} (\bibinfo {year} {2017})},\ \Eprint {http://arxiv.org/abs/1701.08863}
  {arXiv:1701.08863 [astro-ph.HE]} \BibitemShut {NoStop}%
\bibitem [{\citenamefont {Ajello}\ \emph {et~al.}(2011)\citenamefont {Ajello}
  \emph {et~al.}}]{Ajello:2011dq}%
  \BibitemOpen
  \bibfield  {author} {\bibinfo {author} {\bibfnamefont {M.}~\bibnamefont
  {Ajello}} \emph {et~al.} (\bibinfo {collaboration} {The Fermi LAT}),\ }\href
  {\doibase 10.1103/PhysRevD.84.032007} {\bibfield  {journal} {\bibinfo
  {journal} {Phys. Rev.}\ }\textbf {\bibinfo {volume} {D84}},\ \bibinfo {pages}
  {032007} (\bibinfo {year} {2011})},\ \Eprint {http://arxiv.org/abs/1107.4272}
  {arXiv:1107.4272 [astro-ph.HE]} \BibitemShut {NoStop}%
\bibitem [{\citenamefont {Arina}\ \emph {et~al.}(2017)\citenamefont {Arina},
  \citenamefont {Backović}, \citenamefont {Heisig},\ and\ \citenamefont
  {Lucente}}]{Arina:2017sng}%
  \BibitemOpen
  \bibfield  {author} {\bibinfo {author} {\bibfnamefont {C.}~\bibnamefont
  {Arina}}, \bibinfo {author} {\bibfnamefont {M.}~\bibnamefont {Backović}},
  \bibinfo {author} {\bibfnamefont {J.}~\bibnamefont {Heisig}}, \ and\ \bibinfo
  {author} {\bibfnamefont {M.}~\bibnamefont {Lucente}},\ }\href {\doibase
  10.1103/PhysRevD.96.063010} {\bibfield  {journal} {\bibinfo  {journal} {Phys.
  Rev.}\ }\textbf {\bibinfo {volume} {D96}},\ \bibinfo {pages} {063010}
  (\bibinfo {year} {2017})},\ \Eprint {http://arxiv.org/abs/1703.08087}
  {arXiv:1703.08087 [astro-ph.HE]} \BibitemShut {NoStop}%
\bibitem [{\citenamefont {Nomura}\ and\ \citenamefont
  {Thaler}(2009)}]{Nomura:2008ru}%
  \BibitemOpen
  \bibfield  {author} {\bibinfo {author} {\bibfnamefont {Y.}~\bibnamefont
  {Nomura}}\ and\ \bibinfo {author} {\bibfnamefont {J.}~\bibnamefont
  {Thaler}},\ }\href {\doibase 10.1103/PhysRevD.79.075008} {\bibfield
  {journal} {\bibinfo  {journal} {Phys. Rev.}\ }\textbf {\bibinfo {volume}
  {D79}},\ \bibinfo {pages} {075008} (\bibinfo {year} {2009})},\ \Eprint
  {http://arxiv.org/abs/0810.5397} {arXiv:0810.5397 [hep-ph]} \BibitemShut
  {NoStop}%
\bibitem [{\citenamefont {Profumo}\ \emph {et~al.}(2018)\citenamefont
  {Profumo}, \citenamefont {Queiroz}, \citenamefont {Silk},\ and\ \citenamefont
  {Siqueira}}]{Profumo:2017obk}%
  \BibitemOpen
  \bibfield  {author} {\bibinfo {author} {\bibfnamefont {S.}~\bibnamefont
  {Profumo}}, \bibinfo {author} {\bibfnamefont {F.~S.}\ \bibnamefont
  {Queiroz}}, \bibinfo {author} {\bibfnamefont {J.}~\bibnamefont {Silk}}, \
  and\ \bibinfo {author} {\bibfnamefont {C.}~\bibnamefont {Siqueira}},\ }\href
  {\doibase 10.1088/1475-7516/2018/03/010} {\bibfield  {journal} {\bibinfo
  {journal} {JCAP}\ }\textbf {\bibinfo {volume} {1803}},\ \bibinfo {pages}
  {010} (\bibinfo {year} {2018})},\ \Eprint {http://arxiv.org/abs/1711.03133}
  {arXiv:1711.03133 [hep-ph]} \BibitemShut {NoStop}%
\bibitem [{\citenamefont {Albert}\ \emph {et~al.}(2018)\citenamefont {Albert}
  \emph {et~al.}}]{Astropaper}%
  \BibitemOpen
  \bibfield  {author} {\bibinfo {author} {\bibfnamefont {A.}~\bibnamefont
  {Albert}} \emph {et~al.},\ }\href@noop {} {\bibfield  {journal} {\bibinfo
  {journal} {{ArXiv e-prints}}\ } (\bibinfo {year} {2018})},\ \Eprint
  {http://arxiv.org/abs/1808.05620} {arXiv:1808.05620 [astro-ph]} \BibitemShut
  {NoStop}%
\bibitem [{\citenamefont {{Lundberg}}\ and\ \citenamefont
  {{Edsj{\"o}}}(2004)}]{2004PhRvD..69l3505L}%
  \BibitemOpen
  \bibfield  {author} {\bibinfo {author} {\bibfnamefont {J.}~\bibnamefont
  {{Lundberg}}}\ and\ \bibinfo {author} {\bibfnamefont {J.}~\bibnamefont
  {{Edsj{\"o}}}},\ }\href {\doibase 10.1103/PhysRevD.69.123505} {\bibfield
  {journal} {\bibinfo  {journal} {\prd}\ }\textbf {\bibinfo {volume} {69}},\
  \bibinfo {eid} {123505} (\bibinfo {year} {2004})},\ \Eprint
  {http://arxiv.org/abs/astro-ph/0401113} {astro-ph/0401113} \BibitemShut
  {NoStop}%
\bibitem [{\citenamefont {{Flacke}}\ \emph {et~al.}(2009)\citenamefont
  {{Flacke}}, \citenamefont {{Menon}}, \citenamefont {{Hooper}},\ and\
  \citenamefont {{Freese}}}]{2009arXiv0908.0899F}%
  \BibitemOpen
  \bibfield  {author} {\bibinfo {author} {\bibfnamefont {T.}~\bibnamefont
  {{Flacke}}}, \bibinfo {author} {\bibfnamefont {A.}~\bibnamefont {{Menon}}},
  \bibinfo {author} {\bibfnamefont {D.}~\bibnamefont {{Hooper}}}, \ and\
  \bibinfo {author} {\bibfnamefont {K.}~\bibnamefont {{Freese}}},\ }\href@noop
  {} {\bibfield  {journal} {\bibinfo  {journal} {ArXiv e-prints}\ } (\bibinfo
  {year} {2009})},\ \Eprint {http://arxiv.org/abs/0908.0899} {arXiv:0908.0899
  [hep-ph]} \BibitemShut {NoStop}%
\bibitem [{\citenamefont {Griest}\ and\ \citenamefont
  {Seckel}(1987{\natexlab{a}})}]{Griest:1986yu}%
  \BibitemOpen
  \bibfield  {author} {\bibinfo {author} {\bibfnamefont {K.}~\bibnamefont
  {Griest}}\ and\ \bibinfo {author} {\bibfnamefont {D.}~\bibnamefont
  {Seckel}},\ }\href {\doibase 10.1016/0550-3213(87)90293-8,
  10.1016/0550-3213(88)90409-9} {\bibfield  {journal} {\bibinfo  {journal}
  {Nucl. Phys.}\ }\textbf {\bibinfo {volume} {B283}},\ \bibinfo {pages} {681}
  (\bibinfo {year} {1987}{\natexlab{a}})},\ \bibinfo {note} {[Erratum: Nucl.
  Phys.B296,1034(1988)]}\BibitemShut {NoStop}%
\bibitem [{\citenamefont {Gould}(1988)}]{Gould:1987ww}%
  \BibitemOpen
  \bibfield  {author} {\bibinfo {author} {\bibfnamefont {A.}~\bibnamefont
  {Gould}},\ }\href {\doibase 10.1086/166347} {\bibfield  {journal} {\bibinfo
  {journal} {Astrophys. J.}\ }\textbf {\bibinfo {volume} {328}},\ \bibinfo
  {pages} {919} (\bibinfo {year} {1988})}\BibitemShut {NoStop}%
\bibitem [{\citenamefont {Griest}\ and\ \citenamefont
  {Seckel}(1987{\natexlab{b}})}]{GRIEST1987681}%
  \BibitemOpen
  \bibfield  {author} {\bibinfo {author} {\bibfnamefont {K.}~\bibnamefont
  {Griest}}\ and\ \bibinfo {author} {\bibfnamefont {D.}~\bibnamefont
  {Seckel}},\ }\href {\doibase http://dx.doi.org/10.1016/0550-3213(87)90293-8}
  {\bibfield  {journal} {\bibinfo  {journal} {Nuclear Physics B}\ }\textbf
  {\bibinfo {volume} {283}},\ \bibinfo {pages} {681 } (\bibinfo {year}
  {1987}{\natexlab{b}})}\BibitemShut {NoStop}%
\bibitem [{\citenamefont {Gould}(1987)}]{Gould:1987ir}%
  \BibitemOpen
  \bibfield  {author} {\bibinfo {author} {\bibfnamefont {A.}~\bibnamefont
  {Gould}},\ }\href {\doibase 10.1086/165653} {\bibfield  {journal} {\bibinfo
  {journal} {Astrophys. J.}\ }\textbf {\bibinfo {volume} {321}},\ \bibinfo
  {pages} {571} (\bibinfo {year} {1987})}\BibitemShut {NoStop}%
\bibitem [{\citenamefont {Zentner}(2009)}]{Zentner:2009is}%
  \BibitemOpen
  \bibfield  {author} {\bibinfo {author} {\bibfnamefont {A.~R.}\ \bibnamefont
  {Zentner}},\ }\href {\doibase 10.1103/PhysRevD.80.063501} {\bibfield
  {journal} {\bibinfo  {journal} {Phys. Rev.}\ }\textbf {\bibinfo {volume}
  {D80}},\ \bibinfo {pages} {063501} (\bibinfo {year} {2009})},\ \Eprint
  {http://arxiv.org/abs/0907.3448} {arXiv:0907.3448 [astro-ph.HE]} \BibitemShut
  {NoStop}%
\bibitem [{\citenamefont {Rott}\ \emph {et~al.}(2011)\citenamefont {Rott},
  \citenamefont {Tanaka},\ and\ \citenamefont {Itow}}]{Rott:2011fh}%
  \BibitemOpen
  \bibfield  {author} {\bibinfo {author} {\bibfnamefont {C.}~\bibnamefont
  {Rott}}, \bibinfo {author} {\bibfnamefont {T.}~\bibnamefont {Tanaka}}, \ and\
  \bibinfo {author} {\bibfnamefont {Y.}~\bibnamefont {Itow}},\ }\href {\doibase
  10.1088/1475-7516/2011/09/029} {\bibfield  {journal} {\bibinfo  {journal}
  {JCAP}\ }\textbf {\bibinfo {volume} {1109}},\ \bibinfo {pages} {029}
  (\bibinfo {year} {2011})},\ \Eprint {http://arxiv.org/abs/1107.3182}
  {arXiv:1107.3182 [astro-ph.HE]} \BibitemShut {NoStop}%
\bibitem [{\citenamefont {Gould}(1992)}]{Gould:1991hx}%
  \BibitemOpen
  \bibfield  {author} {\bibinfo {author} {\bibfnamefont {A.}~\bibnamefont
  {Gould}},\ }\href {\doibase 10.1086/171156} {\bibfield  {journal} {\bibinfo
  {journal} {Astrophys. J.}\ }\textbf {\bibinfo {volume} {388}},\ \bibinfo
  {pages} {338} (\bibinfo {year} {1992})}\BibitemShut {NoStop}%
\bibitem [{\citenamefont {Batell}\ \emph {et~al.}(2009)\citenamefont {Batell},
  \citenamefont {Pospelov},\ and\ \citenamefont {Ritz}}]{Batell:2009yf}%
  \BibitemOpen
  \bibfield  {author} {\bibinfo {author} {\bibfnamefont {B.}~\bibnamefont
  {Batell}}, \bibinfo {author} {\bibfnamefont {M.}~\bibnamefont {Pospelov}}, \
  and\ \bibinfo {author} {\bibfnamefont {A.}~\bibnamefont {Ritz}},\ }\href
  {\doibase 10.1103/PhysRevD.79.115008} {\bibfield  {journal} {\bibinfo
  {journal} {Phys. Rev.}\ }\textbf {\bibinfo {volume} {D79}},\ \bibinfo {pages}
  {115008} (\bibinfo {year} {2009})},\ \Eprint {http://arxiv.org/abs/0903.0363}
  {arXiv:0903.0363 [hep-ph]} \BibitemShut {NoStop}%
\bibitem [{\citenamefont {{Tang}}\ \emph {et~al.}(2018)\citenamefont {{Tang}},
  \citenamefont {{Ng}}, \citenamefont {{Linden}}, \citenamefont {{Zhou}},
  \citenamefont {{Beacom}},\ and\ \citenamefont
  {{Peter}}}]{2018arXiv180406846T}%
  \BibitemOpen
  \bibfield  {author} {\bibinfo {author} {\bibfnamefont {Q.-W.}\ \bibnamefont
  {{Tang}}}, \bibinfo {author} {\bibfnamefont {K.~C.~Y.}\ \bibnamefont {{Ng}}},
  \bibinfo {author} {\bibfnamefont {T.}~\bibnamefont {{Linden}}}, \bibinfo
  {author} {\bibfnamefont {B.}~\bibnamefont {{Zhou}}}, \bibinfo {author}
  {\bibfnamefont {J.~F.}\ \bibnamefont {{Beacom}}}, \ and\ \bibinfo {author}
  {\bibfnamefont {A.~H.~G.}\ \bibnamefont {{Peter}}},\ }\href {\doibase
  10.1103/PhysRevD.98.063019} {\bibfield  {journal} {\bibinfo  {journal}
  {\prd}\ }\textbf {\bibinfo {volume} {98}},\ \bibinfo {eid} {063019} (\bibinfo
  {year} {2018})},\ \Eprint {http://arxiv.org/abs/1804.06846} {arXiv:1804.06846
  [astro-ph.HE]} \BibitemShut {NoStop}%
\bibitem [{\citenamefont {Atkins}\ \emph {et~al.}(2004)\citenamefont {Atkins}
  \emph {et~al.}}]{PhysRevD.70.083516}%
  \BibitemOpen
  \bibfield  {author} {\bibinfo {author} {\bibfnamefont {R.}~\bibnamefont
  {Atkins}} \emph {et~al.},\ }\href {\doibase 10.1103/PhysRevD.70.083516}
  {\bibfield  {journal} {\bibinfo  {journal} {Phys. Rev. D}\ }\textbf {\bibinfo
  {volume} {70}},\ \bibinfo {pages} {083516} (\bibinfo {year}
  {2004})}\BibitemShut {NoStop}%
\bibitem [{\citenamefont {{Ajello}}\ \emph {et~al.}(2011)\citenamefont
  {{Ajello}} \emph {et~al.}}]{2011PhRvD..84c2007A}%
  \BibitemOpen
  \bibfield  {author} {\bibinfo {author} {\bibfnamefont {M.}~\bibnamefont
  {{Ajello}}} \emph {et~al.},\ }\href {\doibase 10.1103/PhysRevD.84.032007}
  {\bibfield  {journal} {\bibinfo  {journal} {\prd}\ }\textbf {\bibinfo
  {volume} {84}},\ \bibinfo {eid} {032007} (\bibinfo {year} {2011})},\ \Eprint
  {http://arxiv.org/abs/1107.4272} {arXiv:1107.4272 [astro-ph.HE]} \BibitemShut
  {NoStop}%
\bibitem [{\citenamefont {{Zhou}}\ \emph {et~al.}(2017)\citenamefont {{Zhou}},
  \citenamefont {{Ng}}, \citenamefont {{Beacom}},\ and\ \citenamefont
  {{Peter}}}]{2016arXiv161202420Z}%
  \BibitemOpen
  \bibfield  {author} {\bibinfo {author} {\bibfnamefont {B.}~\bibnamefont
  {{Zhou}}}, \bibinfo {author} {\bibfnamefont {K.~C.~Y.}\ \bibnamefont {{Ng}}},
  \bibinfo {author} {\bibfnamefont {J.~F.}\ \bibnamefont {{Beacom}}}, \ and\
  \bibinfo {author} {\bibfnamefont {A.~H.~G.}\ \bibnamefont {{Peter}}},\ }\href
  {\doibase 10.1103/PhysRevD.96.023015} {\bibfield  {journal} {\bibinfo
  {journal} {Phys. Rev. D}\ }\textbf {\bibinfo {volume} {96}},\ \bibinfo {eid}
  {023015} (\bibinfo {year} {2017})},\ \Eprint
  {http://arxiv.org/abs/1612.02420} {arXiv:1612.02420 [astro-ph.HE]}
  \BibitemShut {NoStop}%
\bibitem [{\citenamefont {Ng}\ \emph {et~al.}(2016)\citenamefont {Ng},
  \citenamefont {Beacom}, \citenamefont {Peter},\ and\ \citenamefont
  {Rott}}]{Ng:2015gya}%
  \BibitemOpen
  \bibfield  {author} {\bibinfo {author} {\bibfnamefont {K.~C.~Y.}\
  \bibnamefont {Ng}}, \bibinfo {author} {\bibfnamefont {J.~F.}\ \bibnamefont
  {Beacom}}, \bibinfo {author} {\bibfnamefont {A.~H.~G.}\ \bibnamefont
  {Peter}}, \ and\ \bibinfo {author} {\bibfnamefont {C.}~\bibnamefont {Rott}},\
  }\href {\doibase 10.1103/PhysRevD.94.023004} {\bibfield  {journal} {\bibinfo
  {journal} {Phys. Rev.}\ }\textbf {\bibinfo {volume} {D94}},\ \bibinfo {pages}
  {023004} (\bibinfo {year} {2016})},\ \Eprint
  {http://arxiv.org/abs/1508.06276} {arXiv:1508.06276 [astro-ph.HE]}
  \BibitemShut {NoStop}%
\bibitem [{\citenamefont {{Linden}}\ \emph {et~al.}(2018)\citenamefont
  {{Linden}}, \citenamefont {{Zhou}}, \citenamefont {{Beacom}}, \citenamefont
  {{Peter}}, \citenamefont {{Ng}},\ and\ \citenamefont
  {{Tang}}}]{2018arXiv180305436L}%
  \BibitemOpen
  \bibfield  {author} {\bibinfo {author} {\bibfnamefont {T.}~\bibnamefont
  {{Linden}}}, \bibinfo {author} {\bibfnamefont {B.}~\bibnamefont {{Zhou}}},
  \bibinfo {author} {\bibfnamefont {J.~F.}\ \bibnamefont {{Beacom}}}, \bibinfo
  {author} {\bibfnamefont {A.~H.~G.}\ \bibnamefont {{Peter}}}, \bibinfo
  {author} {\bibfnamefont {K.~C.~Y.}\ \bibnamefont {{Ng}}}, \ and\ \bibinfo
  {author} {\bibfnamefont {Q.-W.}\ \bibnamefont {{Tang}}},\ }\href {\doibase
  10.1103/PhysRevLett.121.131103} {\bibfield  {journal} {\bibinfo  {journal}
  {Physical Review Letters}\ }\textbf {\bibinfo {volume} {121}},\ \bibinfo
  {eid} {131103} (\bibinfo {year} {2018})},\ \Eprint
  {http://arxiv.org/abs/1803.05436} {arXiv:1803.05436 [astro-ph.HE]}
  \BibitemShut {NoStop}%
\bibitem [{\citenamefont {Abdo}\ \emph {et~al.}(2011)\citenamefont {Abdo} \emph
  {et~al.}}]{0004-637X-734-2-116}%
  \BibitemOpen
  \bibfield  {author} {\bibinfo {author} {\bibfnamefont {A.}~\bibnamefont
  {Abdo}} \emph {et~al.},\ }\href
  {http://stacks.iop.org/0004-637X/734/i=2/a=116} {\bibfield  {journal}
  {\bibinfo  {journal} {Astrophys. J.}\ }\textbf {\bibinfo {volume} {734}},\
  \bibinfo {pages} {116} (\bibinfo {year} {2011})}\BibitemShut {NoStop}%
\bibitem [{\citenamefont {{Orlando}}\ and\ \citenamefont
  {{Strong}}(2008{\natexlab{a}})}]{2008A&A...480..847O}%
  \BibitemOpen
  \bibfield  {author} {\bibinfo {author} {\bibfnamefont {E.}~\bibnamefont
  {{Orlando}}}\ and\ \bibinfo {author} {\bibfnamefont {A.~W.}\ \bibnamefont
  {{Strong}}},\ }\href {\doibase 10.1051/0004-6361:20078817} {\bibfield
  {journal} {\bibinfo  {journal} {Astronomy and Astrophysics}\ }\textbf
  {\bibinfo {volume} {480}},\ \bibinfo {pages} {847} (\bibinfo {year}
  {2008}{\natexlab{a}})},\ \Eprint {http://arxiv.org/abs/0801.2178}
  {arXiv:0801.2178} \BibitemShut {NoStop}%
\bibitem [{\citenamefont {Orlando}\ and\ \citenamefont
  {Strong}(2007)}]{Orlando:2006zs}%
  \BibitemOpen
  \bibfield  {author} {\bibinfo {author} {\bibfnamefont {E.}~\bibnamefont
  {Orlando}}\ and\ \bibinfo {author} {\bibfnamefont {A.}~\bibnamefont
  {Strong}},\ }\bibfield  {booktitle} {\emph {\bibinfo {booktitle} {{The
  Multi-Messenger Approach to High-Energy Gamma-Ray Sources: 3rd Workshop on
  the Nature of Unidentified High-Energy Sources, Barcelona, Spain, 4-7 Jul,
  2006}}},\ }\href {\doibase 10.1007/s10509-007-9457-0} {\bibfield  {journal}
  {\bibinfo  {journal} {Astrophys. Space Sci.}\ }\textbf {\bibinfo {volume}
  {309}},\ \bibinfo {pages} {359} (\bibinfo {year} {2007})},\ \Eprint
  {http://arxiv.org/abs/astro-ph/0607563} {arXiv:astro-ph/0607563 [astro-ph]}
  \BibitemShut {NoStop}%
\bibitem [{\citenamefont {Orlando}\ \emph {et~al.}(2017)\citenamefont
  {Orlando}, \citenamefont {Giglietto}, \citenamefont {Moskalenko},
  \citenamefont {Raino'},\ and\ \citenamefont {Strong}}]{Orlando:2017iyc}%
  \BibitemOpen
  \bibfield  {author} {\bibinfo {author} {\bibfnamefont {E.}~\bibnamefont
  {Orlando}}, \bibinfo {author} {\bibfnamefont {N.}~\bibnamefont {Giglietto}},
  \bibinfo {author} {\bibfnamefont {I.}~\bibnamefont {Moskalenko}}, \bibinfo
  {author} {\bibfnamefont {S.}~\bibnamefont {Raino'}}, \ and\ \bibinfo {author}
  {\bibfnamefont {A.}~\bibnamefont {Strong}},\ }\bibfield  {booktitle} {\emph
  {\bibinfo {booktitle} {{Proceedings, 35th International Cosmic Ray Conference
  (ICRC 2017): Bexco, Busan, Korea, July 12-20, 2017}}},\ }\href@noop {}
  {\bibfield  {journal} {\bibinfo  {journal} {PoS}\ }\textbf {\bibinfo {volume}
  {ICRC2017}},\ \bibinfo {pages} {693} (\bibinfo {year} {2017})},\ \Eprint
  {http://arxiv.org/abs/1712.09745} {arXiv:1712.09745 [astro-ph.HE]}
  \BibitemShut {NoStop}%
\bibitem [{\citenamefont {{Orlando}}\ and\ \citenamefont
  {{Strong}}(2008{\natexlab{b}})}]{2008ICRC....2..505O}%
  \BibitemOpen
  \bibfield  {author} {\bibinfo {author} {\bibfnamefont {E.}~\bibnamefont
  {{Orlando}}}\ and\ \bibinfo {author} {\bibfnamefont {A.~W.}\ \bibnamefont
  {{Strong}}},\ }\href@noop {} {\bibfield  {journal} {\bibinfo  {journal}
  {International Cosmic Ray Conference}\ }\textbf {\bibinfo {volume} {2}},\
  \bibinfo {pages} {505} (\bibinfo {year} {2008}{\natexlab{b}})},\ \Eprint
  {http://arxiv.org/abs/0709.3841} {arXiv:0709.3841} \BibitemShut {NoStop}%
\bibitem [{\citenamefont {McDonald}(1998)}]{McDonald1998}%
  \BibitemOpen
  \bibfield  {author} {\bibinfo {author} {\bibfnamefont {F.~B.}\ \bibnamefont
  {McDonald}},\ }\href {\doibase 10.1023/A:1005052908493} {\bibfield  {journal}
  {\bibinfo  {journal} {Space Science Reviews}\ }\textbf {\bibinfo {volume}
  {83}},\ \bibinfo {pages} {33} (\bibinfo {year} {1998})}\BibitemShut {NoStop}%
\bibitem [{\citenamefont {{Abeysekara}}\ \emph {et~al.}(2017)\citenamefont
  {{Abeysekara}} \emph {et~al.}}]{2017ApJ...843...39A}%
  \BibitemOpen
  \bibfield  {author} {\bibinfo {author} {\bibfnamefont {A.~U.}\ \bibnamefont
  {{Abeysekara}}} \emph {et~al.} (\bibinfo {collaboration} {HAWC
  Collaboration}),\ }\href {\doibase 10.3847/1538-4357/aa7555} {\bibfield
  {journal} {\bibinfo  {journal} {Astrophys. J.}\ }\textbf {\bibinfo {volume}
  {843}},\ \bibinfo {eid} {39} (\bibinfo {year} {2017})},\ \Eprint
  {http://arxiv.org/abs/1701.01778} {arXiv:1701.01778 [astro-ph.HE]}
  \BibitemShut {NoStop}%
\bibitem [{\citenamefont {Abeysekara}\ \emph {et~al.}(2014)\citenamefont
  {Abeysekara} \emph {et~al.}}]{Abeysekara:2013qka}%
  \BibitemOpen
  \bibfield  {author} {\bibinfo {author} {\bibfnamefont {A.~U.}\ \bibnamefont
  {Abeysekara}} \emph {et~al.} (\bibinfo {collaboration} {HAWC
  Collaboration}),\ }\href@noop {} {\bibfield  {journal} {\bibinfo  {journal}
  {Braz. J. Phys.}\ }\textbf {\bibinfo {volume} {44}} (\bibinfo {year}
  {2014})},\ \Eprint {http://arxiv.org/abs/1310.0071} {arXiv:1310.0071
  [astro-ph.HE]} \BibitemShut {NoStop}%
\bibitem [{\citenamefont {Abeysekara}\ \emph
  {et~al.}(2017{\natexlab{a}})\citenamefont {Abeysekara} \emph
  {et~al.}}]{Abeysekara:2017hyn}%
  \BibitemOpen
  \bibfield  {author} {\bibinfo {author} {\bibfnamefont {A.~U.}\ \bibnamefont
  {Abeysekara}} \emph {et~al.},\ }\href {\doibase 10.3847/1538-4357/aa7556}
  {\bibfield  {journal} {\bibinfo  {journal} {Astrophys. J.}\ }\textbf
  {\bibinfo {volume} {843}},\ \bibinfo {pages} {40} (\bibinfo {year}
  {2017}{\natexlab{a}})},\ \Eprint {http://arxiv.org/abs/1702.02992}
  {arXiv:1702.02992 [astro-ph.HE]} \BibitemShut {NoStop}%
\bibitem [{\citenamefont {Alfaro}\ \emph {et~al.}(2017)\citenamefont {Alfaro}
  \emph {et~al.}}]{2017arXiv171000890H}%
  \BibitemOpen
  \bibfield  {author} {\bibinfo {author} {\bibfnamefont {R.}~\bibnamefont
  {Alfaro}} \emph {et~al.} (\bibinfo {collaboration} {HAWC Collaboration}),\
  }\href {\doibase 10.1103/PhysRevD.96.122001} {\bibfield  {journal} {\bibinfo
  {journal} {Phys. Rev. D}\ }\textbf {\bibinfo {volume} {96}},\ \bibinfo
  {pages} {122001} (\bibinfo {year} {2017})}\BibitemShut {NoStop}%
\bibitem [{\citenamefont {Baum}\ \emph {et~al.}(2017)\citenamefont {Baum},
  \citenamefont {Visinelli}, \citenamefont {Freese},\ and\ \citenamefont
  {Stengel}}]{Baum:2016oow}%
  \BibitemOpen
  \bibfield  {author} {\bibinfo {author} {\bibfnamefont {S.}~\bibnamefont
  {Baum}}, \bibinfo {author} {\bibfnamefont {L.}~\bibnamefont {Visinelli}},
  \bibinfo {author} {\bibfnamefont {K.}~\bibnamefont {Freese}}, \ and\ \bibinfo
  {author} {\bibfnamefont {P.}~\bibnamefont {Stengel}},\ }\href {\doibase
  10.1103/PhysRevD.95.043007} {\bibfield  {journal} {\bibinfo  {journal} {Phys.
  Rev.}\ }\textbf {\bibinfo {volume} {D95}},\ \bibinfo {pages} {043007}
  (\bibinfo {year} {2017})},\ \Eprint {http://arxiv.org/abs/1611.09665}
  {arXiv:1611.09665 [astro-ph.CO]} \BibitemShut {NoStop}%
\bibitem [{\citenamefont {{Sommerfeld}}(1931)}]{1931AnP...403..257S}%
  \BibitemOpen
  \bibfield  {author} {\bibinfo {author} {\bibfnamefont {A.}~\bibnamefont
  {{Sommerfeld}}},\ }\href {\doibase 10.1002/andp.19314030302} {\bibfield
  {journal} {\bibinfo  {journal} {Annalen der Physik}\ }\textbf {\bibinfo
  {volume} {403}},\ \bibinfo {pages} {257} (\bibinfo {year}
  {1931})}\BibitemShut {NoStop}%
\bibitem [{\citenamefont {Gondolo}\ \emph {et~al.}(2004)\citenamefont
  {Gondolo}, \citenamefont {Edsjo}, \citenamefont {Ullio}, \citenamefont
  {Bergstrom}, \citenamefont {Schelke},\ and\ \citenamefont
  {Baltz}}]{Gondolo:2004sc}%
  \BibitemOpen
  \bibfield  {author} {\bibinfo {author} {\bibfnamefont {P.}~\bibnamefont
  {Gondolo}}, \bibinfo {author} {\bibfnamefont {J.}~\bibnamefont {Edsjo}},
  \bibinfo {author} {\bibfnamefont {P.}~\bibnamefont {Ullio}}, \bibinfo
  {author} {\bibfnamefont {L.}~\bibnamefont {Bergstrom}}, \bibinfo {author}
  {\bibfnamefont {M.}~\bibnamefont {Schelke}}, \ and\ \bibinfo {author}
  {\bibfnamefont {E.~A.}\ \bibnamefont {Baltz}},\ }\href {\doibase
  10.1088/1475-7516/2004/07/008} {\bibfield  {journal} {\bibinfo  {journal}
  {JCAP}\ }\textbf {\bibinfo {volume} {0407}},\ \bibinfo {pages} {008}
  (\bibinfo {year} {2004})},\ \Eprint {http://arxiv.org/abs/astro-ph/0406204}
  {arXiv:astro-ph/0406204 [astro-ph]} \BibitemShut {NoStop}%
\bibitem [{\citenamefont {Bringmann}\ \emph {et~al.}(2018)\citenamefont
  {Bringmann}, \citenamefont {Edsjö}, \citenamefont {Gondolo}, \citenamefont
  {Ullio},\ and\ \citenamefont {Bergström}}]{Bringmann:2018lay}%
  \BibitemOpen
  \bibfield  {author} {\bibinfo {author} {\bibfnamefont {T.}~\bibnamefont
  {Bringmann}}, \bibinfo {author} {\bibfnamefont {J.}~\bibnamefont {Edsjö}},
  \bibinfo {author} {\bibfnamefont {P.}~\bibnamefont {Gondolo}}, \bibinfo
  {author} {\bibfnamefont {P.}~\bibnamefont {Ullio}}, \ and\ \bibinfo {author}
  {\bibfnamefont {L.}~\bibnamefont {Bergström}},\ }\href {\doibase
  10.1088/1475-7516/2018/07/033} {\bibfield  {journal} {\bibinfo  {journal}
  {JCAP}\ }\textbf {\bibinfo {volume} {7}},\ \bibinfo {eid} {033} (\bibinfo
  {year} {2018})},\ \Eprint {http://arxiv.org/abs/1802.03399} {arXiv:1802.03399
  [hep-ph]} \BibitemShut {NoStop}%
\bibitem [{\citenamefont {Sj{\"o}strand}\ \emph {et~al.}(2015)\citenamefont
  {Sj{\"o}strand}, \citenamefont {Ask}, \citenamefont {Christiansen},
  \citenamefont {Corke}, \citenamefont {Desai}, \citenamefont {Ilten},
  \citenamefont {Mrenna}, \citenamefont {Prestel}, \citenamefont {Rasmussen},\
  and\ \citenamefont {Skands}}]{Sjostrand:2014zea}%
  \BibitemOpen
  \bibfield  {author} {\bibinfo {author} {\bibfnamefont {T.}~\bibnamefont
  {Sj{\"o}strand}}, \bibinfo {author} {\bibfnamefont {S.}~\bibnamefont {Ask}},
  \bibinfo {author} {\bibfnamefont {J.~R.}\ \bibnamefont {Christiansen}},
  \bibinfo {author} {\bibfnamefont {R.}~\bibnamefont {Corke}}, \bibinfo
  {author} {\bibfnamefont {N.}~\bibnamefont {Desai}}, \bibinfo {author}
  {\bibfnamefont {P.}~\bibnamefont {Ilten}}, \bibinfo {author} {\bibfnamefont
  {S.}~\bibnamefont {Mrenna}}, \bibinfo {author} {\bibfnamefont
  {S.}~\bibnamefont {Prestel}}, \bibinfo {author} {\bibfnamefont {C.~O.}\
  \bibnamefont {Rasmussen}}, \ and\ \bibinfo {author} {\bibfnamefont {P.~Z.}\
  \bibnamefont {Skands}},\ }\href {\doibase 10.1016/j.cpc.2015.01.024}
  {\bibfield  {journal} {\bibinfo  {journal} {Comput. Phys. Commun.}\ }\textbf
  {\bibinfo {volume} {191}},\ \bibinfo {pages} {159} (\bibinfo {year}
  {2015})},\ \Eprint {http://arxiv.org/abs/1410.3012} {arXiv:1410.3012
  [hep-ph]} \BibitemShut {NoStop}%
\bibitem [{\citenamefont {Abeysekara}\ \emph
  {et~al.}(2017{\natexlab{b}})\citenamefont {Abeysekara} \emph
  {et~al.}}]{Abeysekara:2017mjj}%
  \BibitemOpen
  \bibfield  {author} {\bibinfo {author} {\bibfnamefont {A.~U.}\ \bibnamefont
  {Abeysekara}} \emph {et~al.},\ }\href {\doibase 10.3847/1538-4357/aa7555}
  {\bibfield  {journal} {\bibinfo  {journal} {\apj}\ }\textbf {\bibinfo
  {volume} {843}},\ \bibinfo {eid} {39} (\bibinfo {year}
  {2017}{\natexlab{b}})},\ \Eprint {http://arxiv.org/abs/1701.01778}
  {arXiv:1701.01778 [astro-ph.HE]} \BibitemShut {NoStop}%
\bibitem [{\citenamefont {Abdullah}\ \emph {et~al.}(2014)\citenamefont
  {Abdullah}, \citenamefont {DiFranzo}, \citenamefont {Rajaraman},
  \citenamefont {Tait}, \citenamefont {Tanedo},\ and\ \citenamefont
  {Wijangco}}]{Abdullah:2014lla}%
  \BibitemOpen
  \bibfield  {author} {\bibinfo {author} {\bibfnamefont {M.}~\bibnamefont
  {Abdullah}}, \bibinfo {author} {\bibfnamefont {A.}~\bibnamefont {DiFranzo}},
  \bibinfo {author} {\bibfnamefont {A.}~\bibnamefont {Rajaraman}}, \bibinfo
  {author} {\bibfnamefont {T.~M.~P.}\ \bibnamefont {Tait}}, \bibinfo {author}
  {\bibfnamefont {P.}~\bibnamefont {Tanedo}}, \ and\ \bibinfo {author}
  {\bibfnamefont {A.~M.}\ \bibnamefont {Wijangco}},\ }\href {\doibase
  10.1103/PhysRevD.90.035004} {\bibfield  {journal} {\bibinfo  {journal} {Phys.
  Rev.}\ }\textbf {\bibinfo {volume} {D90}},\ \bibinfo {pages} {035004}
  (\bibinfo {year} {2014})},\ \Eprint {http://arxiv.org/abs/1404.6528}
  {arXiv:1404.6528 [hep-ph]} \BibitemShut {NoStop}%
\bibitem [{\citenamefont {Rajaraman}\ \emph {et~al.}(2015)\citenamefont
  {Rajaraman}, \citenamefont {Smolinsky},\ and\ \citenamefont
  {Tanedo}}]{Rajaraman:2015xka}%
  \BibitemOpen
  \bibfield  {author} {\bibinfo {author} {\bibfnamefont {A.}~\bibnamefont
  {Rajaraman}}, \bibinfo {author} {\bibfnamefont {J.}~\bibnamefont
  {Smolinsky}}, \ and\ \bibinfo {author} {\bibfnamefont {P.}~\bibnamefont
  {Tanedo}},\ }\href@noop {} {\bibfield  {journal} {\bibinfo  {journal} {ArXiv
  e-prints}\ } (\bibinfo {year} {2015})},\ \Eprint
  {http://arxiv.org/abs/1503.05919} {arXiv:1503.05919 [hep-ph]} \BibitemShut
  {NoStop}%
\bibitem [{\citenamefont {Bell}\ \emph {et~al.}(2017)\citenamefont {Bell},
  \citenamefont {Cai}, \citenamefont {Dent}, \citenamefont {Leane},\ and\
  \citenamefont {Weiler}}]{Bell:2017irk}%
  \BibitemOpen
  \bibfield  {author} {\bibinfo {author} {\bibfnamefont {N.~F.}\ \bibnamefont
  {Bell}}, \bibinfo {author} {\bibfnamefont {Y.}~\bibnamefont {Cai}}, \bibinfo
  {author} {\bibfnamefont {J.~B.}\ \bibnamefont {Dent}}, \bibinfo {author}
  {\bibfnamefont {R.~K.}\ \bibnamefont {Leane}}, \ and\ \bibinfo {author}
  {\bibfnamefont {T.~J.}\ \bibnamefont {Weiler}},\ }\href {\doibase
  10.1103/PhysRevD.96.023011} {\bibfield  {journal} {\bibinfo  {journal} {Phys.
  Rev.}\ }\textbf {\bibinfo {volume} {D96}},\ \bibinfo {pages} {023011}
  (\bibinfo {year} {2017})},\ \Eprint {http://arxiv.org/abs/1705.01105}
  {arXiv:1705.01105 [hep-ph]} \BibitemShut {NoStop}%
\bibitem [{\citenamefont {{Choi}}\ \emph {et~al.}(2014)\citenamefont {{Choi}},
  \citenamefont {{Rott}},\ and\ \citenamefont {{Itow}}}]{2014JCAP...05..049C}%
  \BibitemOpen
  \bibfield  {author} {\bibinfo {author} {\bibfnamefont {K.}~\bibnamefont
  {{Choi}}}, \bibinfo {author} {\bibfnamefont {C.}~\bibnamefont {{Rott}}}, \
  and\ \bibinfo {author} {\bibfnamefont {Y.}~\bibnamefont {{Itow}}},\ }\href
  {\doibase 10.1088/1475-7516/2014/05/049} {\bibfield  {journal} {\bibinfo
  {journal} {Journal of Cosmology and Astroparticle Physics}\ }\textbf
  {\bibinfo {volume} {5}},\ \bibinfo {eid} {049} (\bibinfo {year} {2014})},\
  \Eprint {http://arxiv.org/abs/1312.0273} {arXiv:1312.0273 [astro-ph.HE]}
  \BibitemShut {NoStop}%
\bibitem [{\citenamefont {Griest}\ and\ \citenamefont
  {Kamionkowski}(1990)}]{PhysRevLett.64.615}%
  \BibitemOpen
  \bibfield  {author} {\bibinfo {author} {\bibfnamefont {K.}~\bibnamefont
  {Griest}}\ and\ \bibinfo {author} {\bibfnamefont {M.}~\bibnamefont
  {Kamionkowski}},\ }\href {\doibase 10.1103/PhysRevLett.64.615} {\bibfield
  {journal} {\bibinfo  {journal} {Phys. Rev. Lett.}\ }\textbf {\bibinfo
  {volume} {64}},\ \bibinfo {pages} {615} (\bibinfo {year} {1990})}\BibitemShut
  {NoStop}%
\bibitem [{\citenamefont {Blum}\ \emph {et~al.}(2015)\citenamefont {Blum},
  \citenamefont {Cui},\ and\ \citenamefont {Kamionkowski}}]{Blum:2014dca}%
  \BibitemOpen
  \bibfield  {author} {\bibinfo {author} {\bibfnamefont {K.}~\bibnamefont
  {Blum}}, \bibinfo {author} {\bibfnamefont {Y.}~\bibnamefont {Cui}}, \ and\
  \bibinfo {author} {\bibfnamefont {M.}~\bibnamefont {Kamionkowski}},\ }\href
  {\doibase 10.1103/PhysRevD.92.023528} {\bibfield  {journal} {\bibinfo
  {journal} {Phys. Rev.}\ }\textbf {\bibinfo {volume} {D92}},\ \bibinfo {pages}
  {023528} (\bibinfo {year} {2015})},\ \Eprint {http://arxiv.org/abs/1412.3463}
  {arXiv:1412.3463 [hep-ph]} \BibitemShut {NoStop}%
\bibitem [{\citenamefont {von Harling}\ and\ \citenamefont
  {Petraki}(2014)}]{vonHarling:2014kha}%
  \BibitemOpen
  \bibfield  {author} {\bibinfo {author} {\bibfnamefont {B.}~\bibnamefont {von
  Harling}}\ and\ \bibinfo {author} {\bibfnamefont {K.}~\bibnamefont
  {Petraki}},\ }\href {\doibase 10.1088/1475-7516/2014/12/033} {\bibfield
  {journal} {\bibinfo  {journal} {JCAP}\ }\textbf {\bibinfo {volume} {1412}},\
  \bibinfo {pages} {033} (\bibinfo {year} {2014})},\ \Eprint
  {http://arxiv.org/abs/1407.7874} {arXiv:1407.7874 [hep-ph]} \BibitemShut
  {NoStop}%
\bibitem [{\citenamefont {Harigaya}\ \emph {et~al.}(2016)\citenamefont
  {Harigaya}, \citenamefont {Ibe}, \citenamefont {Kaneta}, \citenamefont
  {Nakano},\ and\ \citenamefont {Suzuki}}]{Harigaya:2016nlg}%
  \BibitemOpen
  \bibfield  {author} {\bibinfo {author} {\bibfnamefont {K.}~\bibnamefont
  {Harigaya}}, \bibinfo {author} {\bibfnamefont {M.}~\bibnamefont {Ibe}},
  \bibinfo {author} {\bibfnamefont {K.}~\bibnamefont {Kaneta}}, \bibinfo
  {author} {\bibfnamefont {W.}~\bibnamefont {Nakano}}, \ and\ \bibinfo {author}
  {\bibfnamefont {M.}~\bibnamefont {Suzuki}},\ }\href {\doibase
  10.1007/JHEP08(2016)151} {\bibfield  {journal} {\bibinfo  {journal} {JHEP}\
  }\textbf {\bibinfo {volume} {08}},\ \bibinfo {pages} {151} (\bibinfo {year}
  {2016})},\ \Eprint {http://arxiv.org/abs/1606.00159} {arXiv:1606.00159
  [hep-ph]} \BibitemShut {NoStop}%
\bibitem [{\citenamefont {Bramante}\ and\ \citenamefont
  {Unwin}(2017)}]{Bramante:2017obj}%
  \BibitemOpen
  \bibfield  {author} {\bibinfo {author} {\bibfnamefont {J.}~\bibnamefont
  {Bramante}}\ and\ \bibinfo {author} {\bibfnamefont {J.}~\bibnamefont
  {Unwin}},\ }\href {\doibase 10.1007/JHEP02(2017)119} {\bibfield  {journal}
  {\bibinfo  {journal} {JHEP}\ }\textbf {\bibinfo {volume} {02}},\ \bibinfo
  {pages} {119} (\bibinfo {year} {2017})},\ \Eprint
  {http://arxiv.org/abs/1701.05859} {arXiv:1701.05859 [hep-ph]} \BibitemShut
  {NoStop}%
\bibitem [{\citenamefont {Baldes}\ and\ \citenamefont
  {Petraki}(2017)}]{Baldes:2017gzw}%
  \BibitemOpen
  \bibfield  {author} {\bibinfo {author} {\bibfnamefont {I.}~\bibnamefont
  {Baldes}}\ and\ \bibinfo {author} {\bibfnamefont {K.}~\bibnamefont
  {Petraki}},\ }\href {\doibase 10.1088/1475-7516/2017/09/028} {\bibfield
  {journal} {\bibinfo  {journal} {JCAP}\ }\textbf {\bibinfo {volume} {1709}},\
  \bibinfo {pages} {028} (\bibinfo {year} {2017})},\ \Eprint
  {http://arxiv.org/abs/1703.00478} {arXiv:1703.00478 [hep-ph]} \BibitemShut
  {NoStop}%
\bibitem [{\citenamefont {{Ng}}\ \emph {et~al.}(2017)\citenamefont {{Ng}},
  \citenamefont {{Beacom}}, \citenamefont {{Peter}},\ and\ \citenamefont
  {{Rott}}}]{2017PhRvD..96j3006N}%
  \BibitemOpen
  \bibfield  {author} {\bibinfo {author} {\bibfnamefont {K.~C.~Y.}\
  \bibnamefont {{Ng}}}, \bibinfo {author} {\bibfnamefont {J.~F.}\ \bibnamefont
  {{Beacom}}}, \bibinfo {author} {\bibfnamefont {A.~H.~G.}\ \bibnamefont
  {{Peter}}}, \ and\ \bibinfo {author} {\bibfnamefont {C.}~\bibnamefont
  {{Rott}}},\ }\href {\doibase 10.1103/PhysRevD.96.103006} {\bibfield
  {journal} {\bibinfo  {journal} {\prd}\ }\textbf {\bibinfo {volume} {96}},\
  \bibinfo {eid} {103006} (\bibinfo {year} {2017})},\ \Eprint
  {http://arxiv.org/abs/1703.10280} {arXiv:1703.10280 [astro-ph.HE]}
  \BibitemShut {NoStop}%
\bibitem [{\citenamefont {{Arg{\"u}elles}}\ \emph {et~al.}(2017)\citenamefont
  {{Arg{\"u}elles}}, \citenamefont {{de Wasseige}}, \citenamefont
  {{Fedynitch}},\ and\ \citenamefont {{Jones}}}]{2017JCAP...07..024A}%
  \BibitemOpen
  \bibfield  {author} {\bibinfo {author} {\bibfnamefont {C.~A.}\ \bibnamefont
  {{Arg{\"u}elles}}}, \bibinfo {author} {\bibfnamefont {G.}~\bibnamefont {{de
  Wasseige}}}, \bibinfo {author} {\bibfnamefont {A.}~\bibnamefont
  {{Fedynitch}}}, \ and\ \bibinfo {author} {\bibfnamefont {B.~J.~P.}\
  \bibnamefont {{Jones}}},\ }\href {\doibase 10.1088/1475-7516/2017/07/024}
  {\bibfield  {journal} {\bibinfo  {journal} {Journal of Cosmology and
  Astroparticle Physics}\ }\textbf {\bibinfo {volume} {7}},\ \bibinfo {eid}
  {024} (\bibinfo {year} {2017})},\ \Eprint {http://arxiv.org/abs/1703.07798}
  {arXiv:1703.07798 [astro-ph.HE]} \BibitemShut {NoStop}%
\bibitem [{\citenamefont {Edsjo}\ \emph {et~al.}(2017)\citenamefont {Edsjo},
  \citenamefont {Elevant}, \citenamefont {Enberg},\ and\ \citenamefont
  {Niblaeus}}]{Edsjo:2017kjk}%
  \BibitemOpen
  \bibfield  {author} {\bibinfo {author} {\bibfnamefont {J.}~\bibnamefont
  {Edsjo}}, \bibinfo {author} {\bibfnamefont {J.}~\bibnamefont {Elevant}},
  \bibinfo {author} {\bibfnamefont {R.}~\bibnamefont {Enberg}}, \ and\ \bibinfo
  {author} {\bibfnamefont {C.}~\bibnamefont {Niblaeus}},\ }\href {\doibase
  10.1088/1475-7516/2017/06/033} {\bibfield  {journal} {\bibinfo  {journal}
  {JCAP}\ }\textbf {\bibinfo {volume} {1706}},\ \bibinfo {pages} {033}
  (\bibinfo {year} {2017})},\ \Eprint {http://arxiv.org/abs/1704.02892}
  {arXiv:1704.02892 [astro-ph.HE]} \BibitemShut {NoStop}%
\bibitem [{\citenamefont {{Masip}}(2018)}]{2018APh....97...63M}%
  \BibitemOpen
  \bibfield  {author} {\bibinfo {author} {\bibfnamefont {M.}~\bibnamefont
  {{Masip}}},\ }\href {\doibase 10.1016/j.astropartphys.2017.11.003} {\bibfield
   {journal} {\bibinfo  {journal} {Astroparticle Physics}\ }\textbf {\bibinfo
  {volume} {97}},\ \bibinfo {pages} {63} (\bibinfo {year} {2018})},\ \Eprint
  {http://arxiv.org/abs/1706.01290} {arXiv:1706.01290 [hep-ph]} \BibitemShut
  {NoStop}%
\bibitem [{\citenamefont {In}\ and\ \citenamefont {Rott}(2017)}]{icecubesolar}%
  \BibitemOpen
  \bibfield  {author} {\bibinfo {author} {\bibfnamefont {S.}~\bibnamefont
  {In}}\ and\ \bibinfo {author} {\bibfnamefont {C.}~\bibnamefont {Rott}}
  (\bibinfo {collaboration} {IceCube}),\ }in\ \href
  {https://pos.sissa.it/301/965/pdf} {\emph {\bibinfo {booktitle}
  {{Proceedings, 35th International Cosmic Ray Conference (ICRC 2017): Bexco,
  Busan, Korea, July 12-20, 2017}}}}\ (\bibinfo {year} {2017})\ \Eprint
  {http://arxiv.org/abs/1710.01194} {arXiv:1710.01194} \BibitemShut {NoStop}%
\bibitem [{\citenamefont {Moskalenko}\ and\ \citenamefont
  {Karakula}(1993)}]{0954-3899-19-9-019}%
  \BibitemOpen
  \bibfield  {author} {\bibinfo {author} {\bibfnamefont {I.~V.}\ \bibnamefont
  {Moskalenko}}\ and\ \bibinfo {author} {\bibfnamefont {S.}~\bibnamefont
  {Karakula}},\ }\href {http://stacks.iop.org/0954-3899/19/i=9/a=019}
  {\bibfield  {journal} {\bibinfo  {journal} {Journal of Physics G: Nuclear and
  Particle Physics}\ }\textbf {\bibinfo {volume} {19}},\ \bibinfo {pages}
  {1399} (\bibinfo {year} {1993})}\BibitemShut {NoStop}%
\bibitem [{\citenamefont {Ingelman}\ and\ \citenamefont
  {Thunman}(1996)}]{Ingelman:1996mj}%
  \BibitemOpen
  \bibfield  {author} {\bibinfo {author} {\bibfnamefont {G.}~\bibnamefont
  {Ingelman}}\ and\ \bibinfo {author} {\bibfnamefont {M.}~\bibnamefont
  {Thunman}},\ }\href {\doibase 10.1103/PhysRevD.54.4385} {\bibfield  {journal}
  {\bibinfo  {journal} {Phys. Rev.}\ }\textbf {\bibinfo {volume} {D54}},\
  \bibinfo {pages} {4385} (\bibinfo {year} {1996})},\ \Eprint
  {http://arxiv.org/abs/hep-ph/9604288} {arXiv:hep-ph/9604288 [hep-ph]}
  \BibitemShut {NoStop}%
\bibitem [{\citenamefont {{Rott}}\ \emph {et~al.}(2011)\citenamefont {{Rott}},
  \citenamefont {{Tanaka}},\ and\ \citenamefont
  {{Itow}}}]{2011JCAP...09..029R}%
  \BibitemOpen
  \bibfield  {author} {\bibinfo {author} {\bibfnamefont {C.}~\bibnamefont
  {{Rott}}}, \bibinfo {author} {\bibfnamefont {T.}~\bibnamefont {{Tanaka}}}, \
  and\ \bibinfo {author} {\bibfnamefont {Y.}~\bibnamefont {{Itow}}},\ }\href
  {\doibase 10.1088/1475-7516/2011/09/029} {\bibfield  {journal} {\bibinfo
  {journal} {Journal of Cosmology and Astroparticle Physics}\ }\textbf
  {\bibinfo {volume} {9}},\ \bibinfo {eid} {029} (\bibinfo {year} {2011})},\
  \Eprint {http://arxiv.org/abs/1107.3182} {arXiv:1107.3182 [astro-ph.HE]}
  \BibitemShut {NoStop}%
\bibitem [{\citenamefont {Gao}\ \emph {et~al.}(2018)\citenamefont {Gao},
  \citenamefont {Chen}, \citenamefont {Li}, \citenamefont {Yu}, \citenamefont
  {Liu},\ and\ \citenamefont {He}}]{Gao:2017bfv}%
  \BibitemOpen
  \bibfield  {author} {\bibinfo {author} {\bibfnamefont {B.}~\bibnamefont
  {Gao}}, \bibinfo {author} {\bibfnamefont {S.}~\bibnamefont {Chen}}, \bibinfo
  {author} {\bibfnamefont {Z.}~\bibnamefont {Li}}, \bibinfo {author}
  {\bibfnamefont {C.}~\bibnamefont {Yu}}, \bibinfo {author} {\bibfnamefont
  {K.}~\bibnamefont {Liu}}, \ and\ \bibinfo {author} {\bibfnamefont
  {H.}~\bibnamefont {He}},\ }\bibfield  {booktitle} {\emph {\bibinfo
  {booktitle} {{The Fluorescence detector Array of Single-pixel Telescopes:
  Contributions to the 35th International Cosmic Ray Conference (ICRC
  2017)}}},\ }\href {\doibase 10.22323/1.301.0878} {\bibfield  {journal}
  {\bibinfo  {journal} {PoS}\ }\textbf {\bibinfo {volume} {ICRC2017}},\
  \bibinfo {pages} {878} (\bibinfo {year} {2018})}\BibitemShut {NoStop}%
\bibitem [{\citenamefont {Abercrombie}\ \emph {et~al.}(2015)\citenamefont
  {Abercrombie} \emph {et~al.}}]{Abercrombie:2015wmb}%
  \BibitemOpen
  \bibfield  {author} {\bibinfo {author} {\bibfnamefont {D.}~\bibnamefont
  {Abercrombie}} \emph {et~al.},\ }\href@noop {} {\bibfield  {journal}
  {\bibinfo  {journal} {ArXiv e-prints}\ } (\bibinfo {year} {2015})},\ \Eprint
  {http://arxiv.org/abs/1507.00966} {arXiv:1507.00966 [hep-ex]} \BibitemShut
  {NoStop}%
\bibitem [{\citenamefont {Albert}\ \emph {et~al.}(2017)\citenamefont {Albert}
  \emph {et~al.}}]{Albert:2017onk}%
  \BibitemOpen
  \bibfield  {author} {\bibinfo {author} {\bibfnamefont {A.}~\bibnamefont
  {Albert}} \emph {et~al.},\ }\href@noop {} {\bibfield  {journal} {\bibinfo
  {journal} {ArXiv e-prints}\ } (\bibinfo {year} {2017})},\ \Eprint
  {http://arxiv.org/abs/1703.05703} {arXiv:1703.05703 [hep-ex]} \BibitemShut
  {NoStop}%
\bibitem [{\citenamefont {Aaboud}\ \emph {et~al.}(2018)\citenamefont {Aaboud}
  \emph {et~al.}}]{Aaboud:2017phn}%
  \BibitemOpen
  \bibfield  {author} {\bibinfo {author} {\bibfnamefont {M.}~\bibnamefont
  {Aaboud}} \emph {et~al.} (\bibinfo {collaboration} {ATLAS}),\ }\href
  {\doibase 10.1007/JHEP01(2018)126} {\bibfield  {journal} {\bibinfo  {journal}
  {JHEP}\ }\textbf {\bibinfo {volume} {01}},\ \bibinfo {pages} {126} (\bibinfo
  {year} {2018})},\ \Eprint {http://arxiv.org/abs/1711.03301} {arXiv:1711.03301
  [hep-ex]} \BibitemShut {NoStop}%
\bibitem [{\citenamefont {Sirunyan}\ \emph {et~al.}(2017)\citenamefont
  {Sirunyan} \emph {et~al.}}]{Sirunyan:2017hci}%
  \BibitemOpen
  \bibfield  {author} {\bibinfo {author} {\bibfnamefont {A.~M.}\ \bibnamefont
  {Sirunyan}} \emph {et~al.} (\bibinfo {collaboration} {CMS}),\ }\href
  {\doibase 10.1007/JHEP07(2017)014} {\bibfield  {journal} {\bibinfo  {journal}
  {JHEP}\ }\textbf {\bibinfo {volume} {07}},\ \bibinfo {pages} {014} (\bibinfo
  {year} {2017})},\ \Eprint {http://arxiv.org/abs/1703.01651} {arXiv:1703.01651
  [hep-ex]} \BibitemShut {NoStop}%
\end{thebibliography}%
